\documentclass[traditabstract]{aa} 
\usepackage{graphicx}
\def\la{\lower.5ex\hbox{$\; \buildrel < \over \sim \;$}}
\def\ga{\lower.5ex\hbox{$\; \buildrel > \over \sim \;$}}
\begin{document}
   \title{Large-scale radio continuum properties of 19 Virgo cluster galaxies\thanks{Based on NRAO VLA data (large proposal AV314)}}

   \subtitle{The influence of tidal interactions, ram pressure stripping, and accreting gas envelopes}

   \author{B.~Vollmer\inst{1}, M.~Soida\inst{2}, R.~Beck\inst{3}, A.~Chung\inst{4}, 
     M.~Urbanik\inst{2}, K.T.~Chy\.zy\inst{2}, K.~Otmianowska-Mazur\inst{2}, \and J.D.P.~Kenney\inst{5}}

   \institute{CDS, Observatoire astronomique de Strasbourg, UMR7550, 11, rue de l'universit\'e,
          67000 Strasbourg, France \and
          Astronomical Observatory, Jagiellonian University, Krak\'ow, Poland \and
	  Max-Planck-Institut f\"{u}r Radioastronomie, Auf dem H\"{u}gel 69, 53121 Bonn, Germany \and
          Department of Astronomy and Yonsei University Observatory, Yonsei University, Seoul 120-749, Republic of Korea  \and
	  Yale University Astronomy Department, P.O. Box 208101, New Haven, CT 06520-8101, USA}

   \date{Received ; accepted }


  \abstract
{
Deep scaled array VLA 20 and 6 cm observations including polarization of 19 Virgo spiral galaxies are presented extending
previous work on the influence of the cluster environment on the radio continuum properties of Virgo cluster spiral galaxies.
This sample contains six galaxies with a global minimum of 20~cm polarized emission at the receding side of the galactic disk and quadrupolar type
large-scale magnetic fields.
In the new sample no additional case of a ram-pressure stripped spiral galaxy with an asymmetric ridge of polarized radio continuum
emission was found. In the absence of a close companion, a truncated H{\sc i} disk, together with a ridge of polarized radio continuum 
emission at the outer edge of the H{\sc i} disk,  is a signpost of ram pressure stripping.
Six out of the 19 observed galaxies (NGC~4294, NGC~4298, NGC~4457, NGC~4532, NGC~4568, NGC~4808) display asymmetric 6~cm polarized emission distributions.
Three galaxies belong to tidally interacting pairs (NGC~4294, NGC~4298, NGC~4568), two galaxies host huge accreting H{\sc i} envelopes (NGC~4532, NGC~4808),
and one galaxy (NGC~4457) had a recent minor merger.
Tidal interactions and accreting gas envelopes can lead to compression and shear motions which enhance the polarized radio continuum emission.
The resulting asymmetries are located within the H{\sc i} distribution.
The 6~cm average degree of polarization only correlates with the galaxy mass or rotation velocity. In addition, galaxies with low average star
formation rate per unit area have a low average degree of polarization. Shear or compression motions can enhance the degree of polarization.
The average degree of polarization of tidally interacting galaxies is generally lower than expected for a given rotation velocity and
star formation activity. This low average degree of polarization is at least partly due to the absence of polarized emission from the thin disk.
Ram pressure stripping can decrease whereas tidal interactions most frequently decreases the average degree of 
polarization of Virgo cluster spiral galaxies. We found that moderate active ram pressure stripping has no influence on the spectral index, but
enhances the global radio continuum emission with respect to the FIR emission, 
while an accreting gas envelope can but not necessarily enhances the radio continuum emission with respect to the FIR emission.
}


   \authorrunning{Vollmer et al.}

   \maketitle
%

\section{Introduction\label{sec:introduction}}

Spiral galaxies residing in clusters of galaxies are found to be redder, having less star formation than field galaxy of 
similar Hubble types (Kennicutt 1983, Gavazzi et al. 2006). Since morphology is closely related to the star formation history of a galaxy, 
it is not surprising that the average galaxy properties related to star formation also depend on local 
density (Hashimoto et al. 1998, Lewis et al. 2002, Gomez et al. 2003, Kauffmann et al. 2004, Balogh et al. 2004).
In addition, cluster spiral galaxies are often found to be H{\sc i}-deficient, having lost up to $90$\,\% of their atomic
gas (e.g., Chamaraux et al. 1980, Giovanelli \& Haynes 1983, Cayatte et al. 1994, Chung et al. 2009).
However, their molecular gas content seems not to be different from that of field spiral galaxies 
(Stark et al. 1986, Kenney \& Young 1986, Casoli et al. 1991, Boselli et al. 1995, 2002).
The cluster environment thus alters the properties of spiral galaxies via interactions with the cluster potential,
the intracluster gas (ram pressure stripping), or galaxy-galaxy interactions.
While tidal interactions affect the stars and the interstellar medium of spiral galaxies, ram pressure stripping
only acts on the gas of a galaxy that rapidly moves within the hot cluster atmosphere.

A complementary diagnostic tool for study of interactions of spiral galaxies with the cluster environment are
radio continuum observations. The total radio continuum emission is sensitive to star formation which gives rise to the FIR/radio correlation 
(see, e.g. de Jong et al. 1985, Helou et al. 1985, Condon et al. 1991, Murphy et al. 2008). 
The correlation holds over  5 orders of magnitude and is one of the tightest in astronomy.
The standard deviation of the logarithmic FIR/radio flux ratio $q$ (Helou et al. 1985) is only $0.2$~dex.
Based on 1.4~GHz (20~cm) radio continuum observations Gavazzi et al. (1991) showed that the FIR/radio correlation that is shared by spiral galaxies in a huge 
luminosity interval, is different for cluster and isolated galaxies: the cluster galaxies have an increased radio/FIR ratio. A similar increase 
has been observed at 10.55~GHz (2.8~cm; Niklas et al. 1995) for 6 out of 45 observed Virgo spirals including NGC 4388 and NGC 4438.
Reddy \& Yun (2004) studied the radio and far-infrared (FIR) emission from 114 galaxies in the seven nearest clusters with prominent X-ray emission.
For their luminosity-limited cluster galaxy sample they found an almost 3 times higher scatter in the FIR/radio correlation than in the correlation
for field galaxies. This enhanced scatter is mainly due to radio-excess sources, $\sim 70$\,\% of which show evidence of AGN activity.
The fraction of cluster galaxies with a radio excess is higher in the cluster core than in its periphery, mainly because
radio-luminous early-type galaxies are clustered in the core region.

Miller \& Owen (2001) studied the effect of the cluster environment on the far-infrared FIR/radio correlation in nearby Abell clusters.
They found that the FIR-radio correlation holds quite well for star-forming galaxies. For AGNs, the relative radio-to-FIR fluxes are greater and the scatter in 
the correlation is larger than those seen for star-forming galaxies. Moreover, Miller \& Owen (2001) found some evidence for an excess of 
star-forming galaxies with enhanced radio emission (by a factor of $2$ to $3$) in the centers of the clusters.
This enhancement was not correlated with the line-of-sight velocity relative to the cluster systemic velocities.
Murphy et al. (2009) investigated the radio-FIR relation of Virgo cluster galaxies. They showed that ram pressure affected galaxies 
have global radio flux densities that are enhanced by a factor of 2-3 compared to isolated galaxies.
Whereas the FIR depends on the star formation rate, the radio emission depends on the cosmic ray electron density and the strength of
the galactic magnetic field; both quantities are connected to the star formation rate. 
An enhanced radio/FIR ratio thus implies an enhanced magnetic field strength caused by ISM shear motions or ISM/magnetic field compression by intracluster 
medium ram and/or thermal pressure.

The radio emission of a galaxy can also be normalized to the stellar mass or NIR emission. Ram pressure stripped galaxies
are expected to have truncated gas disks and thus reduced radio/NIR ratios, but they can display enhanced radio/FIR ratios due to
compression of galactic magnetic fields.
Gavazzi \& Boselli (1999) studied the distribution of the radio/NIR luminosity (RLF) of late-type galaxies in 5 nearby galaxy clusters.
They found that the RLF of Cancer, ACO 262 and Virgo are consistent with that of isolated galaxies. However, galaxies in ACO 1367 and 
Coma have their radio emissivity enhanced by a factor $\sim$5 with respect to isolated objects. Multiple systems in the Coma cluster 
bridge also show an enhanced radio/NIR emission. Gavazzi \& Boselli (1999) argue that the latter effect is due to increased star formation 
caused by tidal interactions, whereas the enhanced radio/NIR ratio in ACO 1367 and Coma is due to ram pressure compression of the magnetic 
field.  

Deep VLA imaging observations at 4.86~GHz showed that the distribution of polarized radio continuum emission of 
$8$ Virgo cluster spiral galaxies is strongly asymmetric, with elongated ridges located in the outer galactic disk 
(Vollmer et al. 2004a; Chy\.zy et al. 2006, 2007; Vollmer et al. 2007). These features are not found in similar observations of field galaxies, 
where the distribution of 4.86~GHz polarized emission is generally relatively symmetric and strongest in the interarm regions (Beck 2005). 
The polarized radio continuum emission is sensitive to compression and shear motions within the galactic disks occurring during the interaction 
between the galaxy and its cluster environment. 
Vollmer et al. (2010) compared the radio continuum, H{\sc i}, and H$\alpha$ maps of the $8$ Virgo spiral galaxies and came to the
following conclusions: 
(i) ram pressure compression leads to sharp edges of the H{\sc i} and total power distribution at one side of the galactic disk. 
(ii) The local total power emission is not sensitive to the effects of ram pressure.
(iii) In edge-on galaxies the extraplanar radio emission can extend further than the H{\sc i} emission. 
(iv) In the same galaxies asymmetric gradients in the degree of polarization give additional information on the ram pressure wind direction.  
(v) The radio continuum spectrum might flatten in the compressed region only for very strong ram pressure.

The sample of $8$ Virgo spiral galaxies were carefully selected based on the following criteria: (i) they show signs of interaction with the cluster 
environment such as tidal interactions and/or ram pressure stripping; (ii) VIVA H{\sc i} data (Chung et al. 2009) are available; 
and (iii) their 4.856~GHz total power emission is strong. In previous work we elaborated interaction scenarios for most of the galaxies in our sample. 
To investigate if our conclusions for this sample still hold for a larger sample of Virgo cluster galaxies, we observed another $19$ 
Virgo cluster galaxies of different types and cluster locations.

The galaxies were chosen to be detectable in 4.86~GHz total power emission. 
Flux might have been missed for large target sizes ($R_{25} \sim 5'$). 
These galaxies can be divided into 4 categories according to the following criteria\footnote{We assume a distance of 17~Mpc to the Virgo cluster.}:\\
(i) Galaxy pairs:  NGC~4299/4294, NGC~4298/4302, and NGC~4567/4568.\\
(ii) Galaxies within the cluster Virial radius (distance to M~87 $D \leq 3^{\circ}$($0.9$~Mpc)): 
NGC~4330, NGC~4419, NGC~4424, NGC~457.\\
(iii) Galaxies at the periphery ($4^{\circ} (1.2~{\rm Mpc}) \leq D \leq 8^{\circ}$ (2.4~{\rm Mpc})) of the cluster: 
NGC~4178, NGC~4192, NGC~4216, NGC~4303, NGC4532, NGC4689.\\
(iv) Galaxies far away from the Virgo cluster ($D > 8^{\circ}$): 
NGC~4457, NGC~4713, NGC~4808. 

The present article is structured in the following way: the VLA observations are described in Sect.~\ref{sec:observations}, followed by
a detailed description of the results galaxy per galaxy (Sect.~\ref{sec:results}). The galaxies are then sorted into different classes based on
their radio continuum properties: 20~cm polarized emission asymmetries, symmetric magnetic field patterns, radio continuum halos, asymmetric/truncated
radio continuum halos, asymmetric 6~cm polarized emission distributions (Sect.~\ref{sec:galprops}).
The integrated galaxy properties are discussed in Sect.~\ref{sec:discussion} followed by our conclusions (Sect.~\ref{sec:conclusions}).

\section{Observations\label{sec:observations}}

The 19 Virgo spiral galaxies in 17 VLA fields were observed at 4.86~GHz between October 12, 2009
and December 23, 2009 with the Very Large Array (VLA) of the National
Radio Astronomy Observatory (NRAO)\footnote{NRAO is a facility of
National Science Foundation operated under cooperative agreement by
Associated Universities, Inc.} in the D array configuration. The
band passes were $2\times 50$~MHz. We used 3C286 as the flux
calibrator and 1254+116 as the phase calibrator, the latter of which
was observed every 40~min. Maps were made for both wavelengths using
the AIPS task IMAGR with ROBUST=3, representing a good compromise between resolution and sensitivity.
The final cleaned maps were convolved to a beam size of $22'' \times 22''$ ($1.8 \times 1.8$~kpc). 
In addition, we observed all galaxies at 1.4~GHz on March 21, 2008
in the C array configuration.
The band widths were $2\times 50$~MHz. We used the same calibrators as
for the 4.86~GHz observations. The final cleaned maps were
convolved to a beam size of $22'' \times 22''$.
The rms levels of the 20 and 6~cm total power and
polarized intensity data are shown in Table~\ref{tab:table}.
We obtained apparent B vectors by rotating the observed E vector by $90^{\circ}$,
uncorrected for Faraday rotation\footnote{No corrections for ionospheric Faraday 
rotation were applied.}.
The polarized surface brightness maps were corrected for Ricean bias (using AIPS task COMB with OPCODE='POLC').
The integrated flux densities were calculated by integrating the surface brightness
within the $3\sigma$ r.m.s. contour. For the determination of the mean surface brightness
we divided the integrated flux density by the area of integration.
The given uncertainty does not account for the absolute flux scale uncertainty, but  
the statistical uncertainty based on the r.m.s. and number of independent 
beams used for integration. The degree of polarization was obtained by dividing the
integrated polarized flux density by the integrated total power flux density.
For the spectral index $\alpha$ we used the integrated total power flux densities at 20 and 6~cm:
\begin{equation}
\alpha=\log(S_{\rm TP20}/S_{\rm TP6})/\log(1.49/4.86)\ .
\end{equation}

\begin{figure*}
  \centering
  \resizebox{\hsize}{!}{\includegraphics{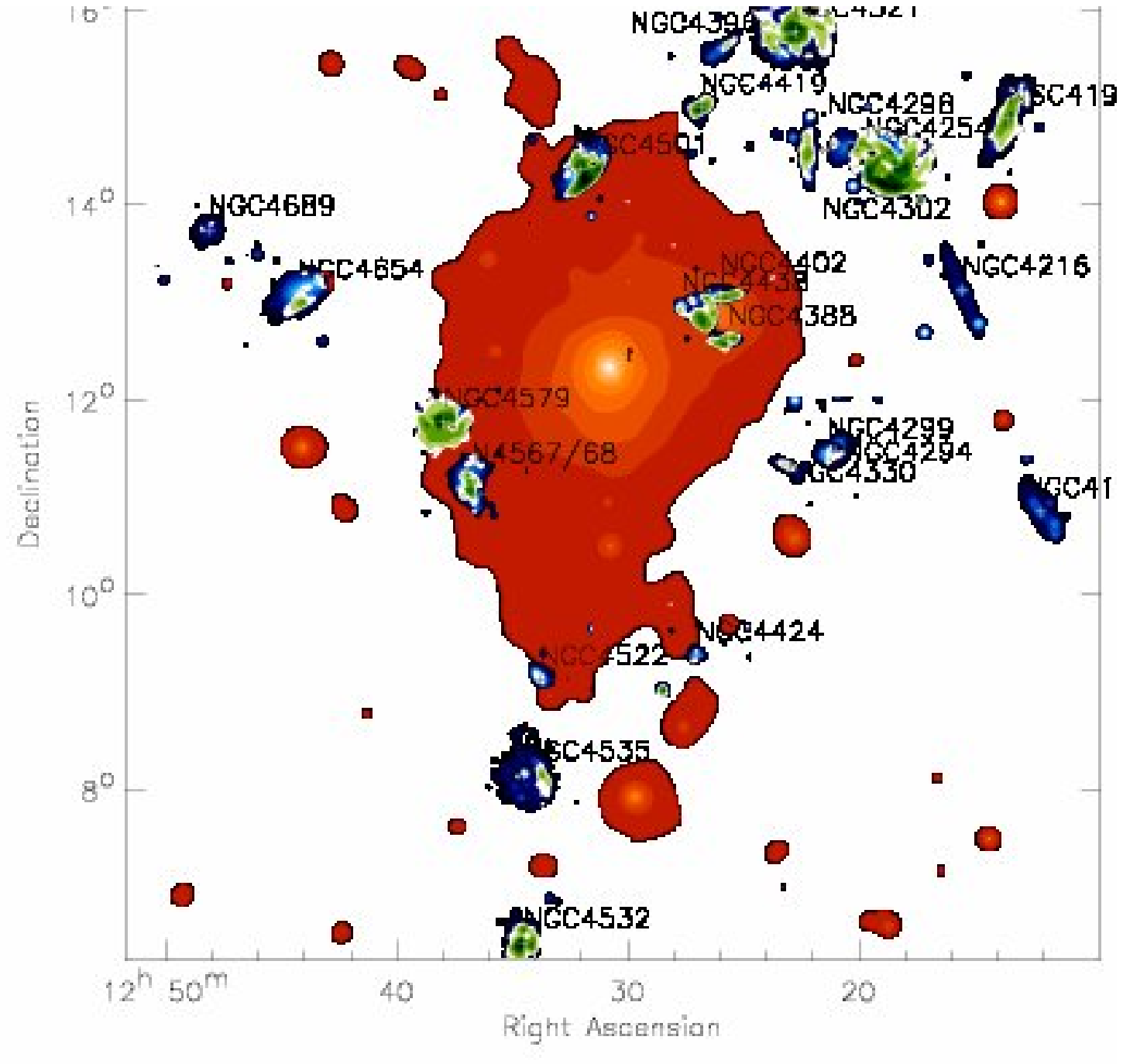}}
  \caption{Virgo cluster galaxies observed in radio continuum emission. Blue: VLA 6~cm total power; green:
VLA 6~cm polarized emission, red: ROSAT intracluster medium X-ray emission (from B\"{o}hringer et al. 1994).
The galaxies are located at the proper position in the cluster but each radio continuum image is magnified by a factor 
10 to show the details of the radio continuum distribution. Radio continuum point sources from the VLA
images were not removed.
The VLA radio continuum data are from this paper, Vollmer et al. (2010), Vollmer et al. (2004a; NGC~4522),
and Chy\.{z}y et al. (2007; NGC~4254). 
  \label{fig:viva}}
\end{figure*}

The integration times and sensitivities of the observations are presented in Table~\ref{tab:table},
the integrated flux densities in Table~\ref{tab:tableflux}, and the average degrees of
polarization in Table~\ref{tab:tableavpol}. An overview of our data is shown in Fig.~\ref{fig:viva}.

\begin{table*}
      \caption{Integration times and rms.}
         \label{tab:table}
      \[
         \begin{array}{llcccccccc}
           \hline
           \noalign{\smallskip}
           {\rm galaxy\ name } & {\rm m_{\rm B}} & i^{(1)} & {\rm Dist.^{(2)}} &{\rm integration}&{\rm integration} &  {\rm rms(TP6cm)} & {\rm rms(TP20cm)} & {\rm rms(PI6cm)} & {\rm rms(PI20cm)}\\
        &  & & & {\rm time\ (6cm)}& {\rm time\ (20cm)}  & (\mu{\rm Jy/} &(\mu{\rm Jy/} & (\mu{\rm Jy/} &(\mu{\rm Jy/}\\
        & {\rm (mag)} & {\rm (deg)}& {\rm (deg)} & {\rm (h:min)} & {\rm (h:min)} & {\rm beam)} & {\rm beam)} & {\rm beam)} & {\rm beam)}\\
       \noalign{\smallskip}
       \hline
       \noalign{\smallskip}
       {\bf NGC~4178} & 11.38 & 70^{\rm (a)} & 4.9 & 6:50 & 1:40  & 9 & 80 & 8 & 20\\
       \noalign{\smallskip}
       \hline
       \noalign{\smallskip}
       {\bf NGC~4192} & 10.95 & 78 & 4.9 & 4:10 & 1:50  & 16 & 210 & 10 & 30\\
       \noalign{\smallskip}
       \hline
       \noalign{\smallskip}
       {\bf NGC~4216} & 10.99 & 85 & 3.8 & 9:20 & 1:40  & 9 & 110 & 7 & 25\\
       \noalign{\smallskip}
       \hline
       \noalign{\smallskip}
       {\bf NGC~4294} & 12.53 & 70 & 2.5 & 6:30 & 1:40  & 10 & 200 & 8 & 30\\
       \noalign{\smallskip}
       \hline
       \noalign{\smallskip}
       {\bf NGC~4298} & 12.04 & 57 & 3.2 & 6:25 & 1:40  & 11 & 170 & 9 & 40\\
       \noalign{\smallskip}
       \hline
       \noalign{\smallskip}
       {\bf NGC~4299} & 12.88 & 22 & 2.5 & 6:20 & 1:40  & 10 & 180 & 7 & 40\\
       \noalign{\smallskip}
       \hline
       \noalign{\smallskip}
       {\bf NGC~4302} & 12.50 & 90 & 3.2 & 6:25 & 1:40  & 11 & 170 & 9 & 40\\
       \noalign{\smallskip}
       \hline
       \noalign{\smallskip}
       {\bf NGC~4303} & 9.97 & 25^{\rm (a)} & 8.2 & 4:50 & 1:35  & 30 & 450 & 10 & 50\\
       \noalign{\smallskip}
       \hline
       \noalign{\smallskip}
       {\bf NGC~4330} & 13.09 & 90 & 2.1 & 6:25 & 1:50  & 14 & 210 & 8 & 40\\
       \noalign{\smallskip}
       \hline
       \noalign{\smallskip}
       {\bf NGC~4419} & 12.08 & 74 & 2.8 & 6:55 & 1:40  & 11 & 130 & 8 & 30\\
       \noalign{\smallskip}
       \hline
       \noalign{\smallskip}
       {\bf NGC~4424} & 12.34 & 62 & 3.1 & 7:00 & {\rm VIVA}  & 14 & - & 7 & - \\
       \noalign{\smallskip}
       \hline
       \noalign{\smallskip}
       {\bf NGC~4457} & 11.76 & 33 & 8.8 & 2:40 & 1:40  & 23 & 190 & 13 & 50\\
       \noalign{\smallskip}
       \hline
       \noalign{\smallskip}
       {\bf NGC~4532} & 12.30 & 70 & 6.0 & 6:25 & 1:40  & 18 & 160 & 8 & 30\\
       \noalign{\smallskip}
       \hline
       \noalign{\smallskip}
       {\bf NGC~4567/68} & 12.06/11.68 & 49/66 & 1.8 & 5:00 & 1:40  & 14 & 230 & 9 & 45\\
       \noalign{\smallskip}
       \hline
       \noalign{\smallskip}
       {\bf NGC~4579} & 10.48 & 38 & 1.8 & 5:40 & 1:55  & 18 & 250 & 8 & 45\\
       \noalign{\smallskip}
       \hline
       \noalign{\smallskip}
       {\bf NGC~4689} & 11.60 & 37 & 4.5 & 3:10 & 1:40  & 14 & 80 & 12 & 20\\
       \noalign{\smallskip}
       \hline
       \noalign{\smallskip}
       {\bf NGC~4713} & 12.19 & 52 & 8.5 & 6:25 & 1:40  & 12 & 120 & 8 & 25\\
       \noalign{\smallskip}
       \hline
       \noalign{\smallskip}
       {\bf NGC~4808} & 12.35 & 68 & 10.2 & 6:25 & 1:40  & 14 & 140 & 9 & 30\\
       \noalign{\smallskip}
       \hline
        \end{array}
      \]
\begin{list}{}{}
\item[$^{(1)}$ Inclination angle; $^{(2)}$ Distance from M~87]
\item[$^{(a)}$ from Cayatte et al. (1990)]
\end{list}
\end{table*}

\section{Results\label{sec:results}}

\begin{table*}
      \caption{Integrated flux densities.}
         \label{tab:tableflux}
      \[
         \begin{array}{lcrrrrrrc}
           \hline
           \noalign{\smallskip}
           {\rm galaxy\ name } & {\rm type}^{(a)} & v_{\rm rot}^{(b)} & {\rm def}_{\rm HI}^{(c)} & S^{(d)}_{\rm TP20cm} & S_{\rm PI20cm} & S^{(d)}_{\rm TP6cm} & S_{\rm PI6cm} & SB^{(de)}_{\rm TP6cm}  \\
	    & & ({\rm km/s}) & & {\rm (mJy)} & {\rm (mJy)} & {\rm (mJy)} & {\rm (mJy)} & (\mu{\rm Jy/beam)}\\
       \noalign{\smallskip}
       \hline
       \noalign{\smallskip}
       {\bf NGC~4178} & {\rm SB(rs)dm} & 145 & -0.23 & 30.7 \pm 1.1 & 0.5 \pm 0.2 & 13.0 \pm 0.3 & 0.5 \pm 0.1 & 154 \\
       \noalign{\smallskip}
       \hline
       \noalign{\smallskip}
       {\bf NGC~4192} & {\rm SAB(s)ab} & 245 & 0.51& 120.9(104.3) \pm 3.3 & 3.0 \pm 0.5 & 35.0(28.7) \pm 0.8 & 4.6 \pm 0.3 & 282(231)\\
       \noalign{\smallskip}
       \hline
       \noalign{\smallskip}
       {\bf NGC~4216} & {\rm SAB(s)b} & 270 & 0.76 &  19.9 \pm 1.3 & 0.8 \pm 0.2 & 11.6(6.5) \pm 0.3 & 0.3 \pm 0.1 & 157(88) \\
       \noalign{\smallskip}
       \hline
       \noalign{\smallskip}
       {\bf NGC~4294} & {\rm SB(s)cd} & 120 & -0.11 &  32.3 \pm 1.7 & 0.4 \pm 0.3 & 11.5 \pm 0.2 & 0.8 \pm 0.1 & 311 \\
       \noalign{\smallskip}
       \hline
       \noalign{\smallskip}
       {\bf NGC~4298} & {\rm SA(rs)c} & 145 & 0.41 &  24.8 \pm 1.4 & 1.3 \pm 0.3 & 8.0 \pm 0.2 & 0.7 \pm 0.1 & 401 \\
       \noalign{\smallskip}
       \hline
       \noalign{\smallskip}
       {\bf NGC~4299} & {\rm SAB(s)dm} &185  & -0.43 &  17.6 \pm 1.0 & 0.0 \pm 0.2 & 7.7 \pm 0.2 & 0.4 \pm 0.1 & 345 \\
       \noalign{\smallskip}
       \hline
       \noalign{\smallskip}
       {\bf NGC~4302} & {\rm Sc} & 190  & 0.39 &  43.2 \pm 1.7 & 3.0 \pm 0.5 & 13.2 \pm 0.2 & 2.0 \pm 0.1 & 526 \\
       \noalign{\smallskip}
       \hline
       \noalign{\smallskip}
       {\bf NGC~4303} & {\rm SAB(rs)bc} & 200 & 0.08 &  421.8(398.5) \pm 7.3 & 9.6 \pm 1.3 & 145.2(135.1) \pm 1.1 & 18.6 \pm 0.4 & 1618(1506) \\
       \noalign{\smallskip}
       \hline
       \noalign{\smallskip}
       {\bf NGC~4330} & {\rm Scd?} & 140 & 0.80 &  13.3 \pm 1.2 & 0.6 \pm 0.3 & 4.7 \pm 0.3 & 0.6 \pm 0.1 & 152 \\
       \noalign{\smallskip}
       \hline
       \noalign{\smallskip}
       {\bf NGC~4419} & {\rm SB(s)a} & 200 & 1.37 &  59.8(17.7) \pm 0.9 & 0.6 \pm 0.3 & 23.6(9.2) \pm 0.3 & 1.4 \pm 0.1 & 915(355) \\
       \noalign{\smallskip}
       \hline
       \noalign{\smallskip}
       {\bf NGC~4424} & {\rm SB(s)a} & 60 & 0.97 & 6.5 \pm 0.7 & - & 4.7 \pm 0.2 & 0.2 \pm 0.1 &  568 \\
       \noalign{\smallskip}
       \hline
       \noalign{\smallskip}
       {\bf NGC~4457} & {\rm (R)SAB(s)0} & 150 & 0.92  & 34.1 \pm 1.0 & 1.1 \pm 0.4 & 16.2 \pm 0.3 & 1.9 \pm 0.1 & 596 \\
       \noalign{\smallskip}
       \hline
       \noalign{\smallskip}
       {\bf NGC~4532} & {\rm IBm} & 110 & -0.06  & 119.9 \pm 1.7 & 1.0 \pm 0.3 & 51.2 \pm 0.5 & 4.2 \pm 0.2 & 772 \\
       \noalign{\smallskip}
        \hline
       \noalign{\smallskip}
       {\bf NGC~4567/68} & {\rm SA(rs)bc} & 135/185 & 0.13/0.38  & 128.8 \pm 1.3 & 2.9 \pm 0.5 & 46.8 \pm 0.2 & 2.6 \pm 0.2 & 1622 \\
       \noalign{\smallskip}
       \hline
       \noalign{\smallskip}
       {\bf NGC~4579} & {\rm SAB(rs)b} & 300 & 0.95  & 133.4(88.4) \pm 2.0 & 8.5 \pm 0.7 & 81.3(43.3) \pm 0.7 & 6.0 \pm 0.2 & 1098(585) \\
       \noalign{\smallskip}
       \hline
       \noalign{\smallskip}
       {\bf NGC~4689} & {\rm SA(rs)bc} & 165 & 0.68  & 19.3 \pm 0.7 & 1.3 \pm 0.3 & 6.5 \pm 0.3 & 0.6 \pm 0.2 & 149 \\
       \noalign{\smallskip}
       \hline
       \noalign{\smallskip}
       {\bf NGC~4713} & {\rm SAB(rs)d} & 120 & -0.31 & 48.7 \pm 1.3 & 2.2 \pm 0.4 & 17.8 \pm 0.2 & 1.8 \pm 0.1 & 461 \\
       \noalign{\smallskip}
       \hline
       \noalign{\smallskip}
       {\bf NGC~4808} & {\rm SA(s)cd} & 150 & -0.58 & 62.6 \pm 2.0 & 2.2 \pm 0.2 & 19.4 \pm 0.2 & 1.1 \pm 0.1 & 778 \\
       \noalign{\smallskip}
       \hline
         \end{array}
	 \]
	 \begin{list}{}{}
	 \item[$^{(a)}$ Morphological type from RC3 (de Vaucouleurs et al. 1991),]
	 \item[$^{(b)}$ rotation velocity based on the HI linewidth (20\%) and inclination,]
	 \item[$^{(c)}$ HI deficiency from Chung et al. (2009) and Cayatte et al. (1994),]
	 \item[$^{(d)}$ a point source has been removed for the value in brackets,]
	 \item[$^{(e)}$ surface brightness within the radio continuum isophotes.]
	 \end{list}
\end{table*}

\begin{table}
      \caption{Average degree of polarization and spectral index.}
         \label{tab:tableavpol}
      \[
         \begin{array}{lccc}
           \hline
           \noalign{\smallskip}
           {\rm galaxy\ name } & $$\Pi_{\rm 20cm}$$\ (\%) & $$\Pi_{\rm 6cm}$$\ (\%) & {\rm SI} \\
       \noalign{\smallskip}
       \hline
       \noalign{\smallskip}
       {\bf NGC~4178} & 1.7\pm 0.7 & 3.5\pm 0.8 & -0.69\pm 0.03 \\
       \noalign{\smallskip}
       \hline
       \noalign{\smallskip}
       {\bf NGC~4192^*} & 2.8\pm 0.5 & 16.0\pm 1.1 & -1.03\pm 0.03 \\
       \noalign{\smallskip}
       \hline
       \noalign{\smallskip}
       {\bf NGC~4216^*} & 3.9\pm 1.0 & 4.3\pm 1.6 & -0.90\pm 0.06 \\
       \noalign{\smallskip}
       \hline
       \noalign{\smallskip}
       {\bf NGC~4294} & 1.1\pm 0.9 & 6.8\pm 0.9 & -0.83\pm 0.04 \\
       \noalign{\smallskip}
       \hline
       \noalign{\smallskip}
       {\bf NGC~4298} & 5.1\pm 1.2 & 8.6\pm 1.3 & -0.91\pm 0.05 \\
       \noalign{\smallskip}
       \hline
       \noalign{\smallskip}
       {\bf NGC~4299} & 0.0\pm 1.1 & 5.5\pm 1.3 & -0.66\pm 0.05 \\
       \noalign{\smallskip}
       \hline
       \noalign{\smallskip}
       {\bf NGC~4302} & 6.9\pm 1.2 & 15.2\pm 0.8 & -0.95\pm 0.03 \\
       \noalign{\smallskip}
       \hline
       \noalign{\smallskip}
       {\bf NGC~4303^*} & 2.4\pm 0.3 & 13.8\pm 0.3 & -0.87\pm 0.02 \\
       \noalign{\smallskip}
       \hline
       \noalign{\smallskip}
       {\bf NGC~4330} & 4.5\pm 2.3 & 13.3\pm 2.3 & -0.84\pm 0.09 \\
       \noalign{\smallskip}
       \hline
       \noalign{\smallskip}
       {\bf NGC~4419^*} & 3.4\pm 1.7 & 15.2\pm 1.2 & -0.53\pm 0.05 \\
       \noalign{\smallskip}
       \hline
       \noalign{\smallskip}
       {\bf NGC~4424} & - & 4.5\pm 2.1 & -0.26\pm 0.09 \\
       \noalign{\smallskip}
       \hline
       \noalign{\smallskip}
       {\bf NGC~4457} & 3.1\pm 1.2 & 12.0\pm 0.7 & -0.60\pm 0.03 \\
       \noalign{\smallskip}
       \hline
       \noalign{\smallskip}
       {\bf NGC~4532} & 0.9\pm 0.3 & 8.2\pm 0.4 & -0.68\pm 0.01\\
       \noalign{\smallskip}
        \hline
       \noalign{\smallskip}
       {\bf NGC~4567/68} & 2.3\pm 0.4 & 5.6\pm 0.4 & -0.81\pm 0.01 \\
       \noalign{\smallskip}
       \hline
       \noalign{\smallskip}
       {\bf NGC~4579^*} & 9.6\pm 0.8 & 13.9\pm 0.5 & -0.57\pm 0.02 \\
       \noalign{\smallskip}
       \hline
       \noalign{\smallskip}
       {\bf NGC~4689} & 6.8\pm 1.6 & 9.7\pm 3.1 & -0.87\pm 0.05 \\
       \noalign{\smallskip}
       \hline
       \noalign{\smallskip}
       {\bf NGC~4713} & 4.4\pm 0.8 & 10.1\pm 0.6 & -0.81\pm 0.02 \\
       \noalign{\smallskip}
       \hline
       \noalign{\smallskip}
       {\bf NGC~4808} & 3.5\pm 0.3 & 5.9\pm 0.5 & -0.94\pm 0.03 \\
       \noalign{\smallskip}
       \hline
         \end{array}
	 \]
	 \begin{list}{}{}
	 \item[$*$: nuclear emission subtracted. ]
	 \end{list}
\end{table}

\subsection{Galaxy properties \label{sec:properties}}

In the following, the radio continuum properties of the observed sample will be described galaxy by galaxy.
To do so, we compare the (i) total power and polarized emission, (ii) the projected magnetic regular field structure, and 
(iii) the spectral index and degree of polarization maps to the optical, H$\alpha$ (Goldmine; Gavazzi et al. 2003), and H{\sc i} 
VIVA, Chung et al. 2009) images (see Appendix A). If available, we give additional information on the galaxy.

\subsubsection{NGC~4178}

This highly inclined galaxy has an asymmetric radio continuum emission distribution at 6 and 20~cm,
the southwestern part being brighter than the northeastern part of the galactic disk. 
This asymmetry along the galaxy's major axis is also observed in H$\alpha$.
The overall radio continuum surface brightness is low ($\sim 200$~$\mu$Jy/beam at 6~cm).
No polarized emission is detected neither at 20~cm nor at 6~cm. The average degree of polarization
is very low ($< 4$\,\%). The spectral index is shallow ($-0.6$) in starforming regions with H$\alpha$ emission.

\subsubsection{NGC~4192}

The overall surface brightness of the radio continuum disk of this highly inclined galaxy is low
($\sim 300$~$\mu$Jy/beam at 6~cm) as the star formation activity (H$\alpha$). 
The 20 and 6~cm radio continuum emission distributions are symmetric. The spectral index is steep
($< -1$) all over the disk. A vertically extended radio continuum halo is observed at 6~cm.
Polarized emission at 20~cm is only detected along the minor axis to the southwest.
Since this is the far side of the galactic disk, this asymmetry can be explained by emission from 
regions above the thin galactic disk (near the observer), the emission from below
being depolarized by the turbulent magnetic field of the thin disk.
At 6~cm the polarized emission extends over the whole galactic disk. 
The degree of polarization is rather high ($\sim 16$\,\%). The regular magnetic field
displays an X structure in the disk which is more pronounced on the southwestern side of galaxy's major axis.
These X structures are frequently observed in the halos of starforming edge-on spiral galaxies (e.g., NGC~253, Heesen et al. 2009;
NGC~5775, Soida et al. 2011).
The X-shaped magnetic field of NGC~4192 is thus may be the projection of the halo field if the regular disk field is weak.

\subsubsection{NGC~4216}

This anemic edge-on galaxies has a very low H{\sc i} column density and 
star formation activity (H$\alpha$). It shows the lowest radio contiuum surface brightness ($\sim 100$~m$\mu$Jy/beam at 6~cm).
The disk emission is barely detected at 20~cm and exceptionally thin at 6~cm.
The spectral index is steep ($\sim -0.9$) everywhere in the disk.
Polarized emission at 20 or 6~cm is not detected in this galaxy. As NGC~4178, the average degree of polarization
is very low ($< 5$\,\%).

\subsubsection{NGC~4294}

NGC~4294 is part of a galaxy pair. Together with NGC~4299, it shares a common H{\sc i} envelope (Chung et al. 2009). 
However, none of these galaxies shows a significantly perturbed optical disk, reminiscent of a tidal interaction.
The radio continuum disk of this highly inclined disk is symmetric at both frequencies. 
It is considerably more extended in the vertical direction at 6~cm than at 20~cm.
The spectral index of $-0.8$ is normal for a small, moderately starforming spiral galaxy. 
No polarized emission is detected at 20~cm. At 6~cm, the polarized emission is highly asymmtric,
with its maximum located in the southeastern part of the galactic disk. The magnetic field structure in
this region does not follow a spiral structure, but has a dominant vertical component.

\subsubsection{NGC~4298}

This galaxy forms a pair with NGC~4302. Its optical and H{\sc i} disks are moderately asymmetric along the major axis
with a more extended disk to the southeast. In addition, the H{\sc i} is more extended and diffuse to the northwest
while the other side of the disk shows a sharp cutoff (Chung et al. 2009). The radio continuum emission at 20 and 6~cm follow this asymmetry.
The 6~cm emission extends beyond the optical main disk and H{\sc i} disk northeast and west from the galaxy center.
The spectral index is normal ($-0.7$) in starforming regions (H$\alpha$) and steep ($-0.9$) elsewhere.
No polarized emission is detected at 20~cm. At 6~cm, polarized emission is only detected northeast of the galaxy center.
In this region the degree of polarization is high ($\sim 25$\,\%).

\subsubsection{NGC~4299}

Together with NGC~4294, NGC~4299 shares a common H{\sc i} envelope (Chung et al. 2009). 
The radio continuum emission of this face-on galaxy is symmetric and more extended at 6~cm than at 20~cm.
The spectral index of the galactic disk is quite shallow ($\sim -0.6$). 
Polarized emission is only detected at 6~cm at a low level. The magnetic field structure seems to be that of a
symmetric spiral where detected. The three distinct regions of polarized 6~cm emission with similar pitch angles of
the magnetic field might be outer magnetic arms.

\subsubsection{NGC~4302}
 
This edge-on galaxy forms a pair with NGC~4298. The H{\sc i} distribution is mildly truncated to within the optical disk
to the south, while a long tail is present in the opposite side (Chung et al. 2009).
The radio continuum disk is symmetric and thin at 20 and 6~cm. The latter emission is radially and vertically more extended
than the 20~cm emission. The spectral index is steep ($-1$) over the whole galactic disk.
At the northern edge of the optical disk a point source
is detected at 6~cm, but not at 20~cm. The most probable explanation for the absence of the source at 20~cm is
a supernova explosion between April 2008 and October 2009 giving rise to a supernova remnant detected in the 
radio continuum. Polarized emission at 20~cm is only detected in the northern half of the galactic disk
where strong Faraday rotation changes the angles of the magnetic fields significantly.
At 6~cm the whole disk is visible in polarized radio continuum emission. The average degree of polarization at 6~cm is $15$\,\%.
As in NGC~4192, the regular magnetic field shows an X structure.

\subsubsection{NGC~4303}

This face-on galaxy displays a symmetric disk in all tracers (radio continuum, optical, H$\alpha$, H{\sc i} (Cayatte et al. 1990)),
comparable to NGC~4321 (Vollmer et al. 2010). It has a high degree of polarization ($15$\,\%), a normal spectral index ($-0.8$)
in starforming regions and a steeper spectral index elsewhere. The magnetic field pattern is that of a symmetric spiral.
However, it is exceptionally smooth at 6~cm, not concentrated in the interarm regions as in many field spiral galaxies,
and does not follow the kinks of the optical spiral arms. The degree of polarization increases with increasing galactic radius.
The average degree of polarization at 6~cm is $14$\,\%.
At 20~cm we observe a strong Faraday depolarization in the inner disk, where polarized emission at 20~cm is not detected.

\subsubsection{NGC~4330}

The edge-on galaxy NGC~4330 has a prominent H{\sc i} and UV tail to the southwest (Chung et al. 2009, Abramson et al. 2011).
The UV and H{\sc i} morphologies are interpreted as signs of ongoing ram pressure stripping (Abramson et al. 2011).
Murphy et al. (2009) found that the local radio/FIR ratio is lower than the global ratio at the northern border of the H{\sc i} disk which they
ascribe to ongoing ram pressure stripping. The 6~cm continuum emission is asymmetric with respect to the major
axis with more emission to the southeast. A radio continuum tail is observed which is offset to the east from the H{\sc i} tail.
The quality of the 20~cm data is not good enough to detect this radio tail. The spectral index and the degree of polarization are normal.
The 6~cm radio continuum data were already presented in Vollmer et al. (2012a) together with a dynamical model including
ram pressure stripping. Vollmer et al. (2012a) concluded that NGC~4330 moves to the north and still approaches the cluster center 
with the closest approach occurring in ~100 Myr. In contrast to other Virgo spiral galaxies affected by ram pressure stripping, 
NGC 4330 does not show an asymmetric ridge of polarized radio continuum emission. Vollmer et al. (2012a) suggested that this is due to the relatively slow 
compression of the ISM and the particular projection of NGC 4330. The magnetic field in the northwestern side of the halo looks normal
with an X-shaped field pattern, while the southern side is heavily distorted by ram pressure.

\subsubsection{NGC~4419}

This highly inclined galaxy is strongly H{\sc i} deficient and the H{\sc i} distribution is severely truncated
within the stellar disk. It has strong radio continuum emission from a nuclear source, likely an AGN (Decarli et al. 2007).
The total gas surface density distribution (CO + H{\sc i}) is strongly 
peaked at the galaxy center (Vollmer et al. 2012b). Due to the high gas surface density, the star formation rate per unit area is also 
high. As the gas distribution, the radio continuum distribution at 20 and 6~cm is strongly truncated. In addition, a vertically extended radio
halo is observed with a characteristic X structure of the magnetic field. The regular magnetic field in the disk is parallel to the disk plane.
Polarized emission at 20~cm is only observed southeast of the
galaxy center. The spectral index and the degree of polarization are normal.

\subsubsection{NGC~4424 \label{sec:n4424}}

NGC~4424 is classified as an H{\sc ii} galaxy. It has a strongly disturbed stellar disk, with banana-shaped isophotes and shells (Cort\'es et al. 2006).
There is strong star formation in a bar-like string of H{\sc ii} complexes in the central 1~kpc, and no star formation beyond (Kenney et al. 1996).
Its stellar rotation velocities ($\sim 30$~km\,s$^{-1}$) are extremely modest. A prominent H{\sc i} tail extends to the south (Chung et al. 2007, 2009).
Cort\'es et al. (2006) suggested that the peculiarities of NGC~4424 are the result of an intermediate-mass merger plus ram pressure stripping. 
The 20 and 6~cm radio continuum emission distribution is almost round. The spectral index is extremely flat ($-0.3$), i.e. the fraction of thermal
to synchrotron emission is high. No polarized emission has been observed at 20~cm. At 6~cm the polarized emission has two lobes to the north and south.
This morphology together with the vertical magnetic field structure stems from a central outflow.

\subsubsection{NGC~4457}

NGC~4457 has an asymmetric H$\alpha$ distribution with a prominent arm in the southern galactic disk.
The H{\sc i} extent is smaller than the optical disk and quite asymmetric with the peak coinciding with the prominent H$\alpha$ arm (Chung et al. 2009).
The radio continuum emission distribution is extended and asymmetric with an arm structure to the east. This arm extends beyond the
H{\sc i} distribution. While the spectral index is 
shallow ($-0.5$) in the starforming regions, it is steep ($-0.9$) in the eastern arm structure.  
Polarized emission at 20~cm is only observed in the arm structure, whereas polarized emission is ubiquitous at 6~cm.
The magnetic field is symmetric with a small pitch angle.

\subsubsection{NGC~4532}

NGC 4532 is an H{\sc i}-rich, optically peculiar Sm galaxy with strong H$\alpha$ emission indicating a high star formation rate.
NGC~4532 shares a common H{\sc i} distribution with DDO~137 (Koopmann et al. 2008). In addition, a $\sim 500$~kpc H{\sc i} extension was discovered by
Koopmann et al. (2008) southwest of the galaxy pair. In particular, an H{\sc i} filament is running in projection across NGC~4532 in east-west 
direction. The 20 and 6~cm continuum emission is vertically extended. The spectral index significantly steepens with increasing height. 
No polarized emission is detected at 20~cm. The polarized emission at 6~cm displays two lobes along the minor axis (east and west).
The galactic disk seems thus to be highly depolarized. The structure of the magnetic field has a dominant vertical direction and can be
seen as an extreme X structure.

\subsubsection{NGC~4567/68}

NGC~4567/68 is a close, overlapping pair of Sc galaxies in the Virgo cluster. 
Their H{\sc i} line-of-sight velocities match where the galaxies overlap (Chung et al. 2009).
Optically, neither shows significant disturbances in their inner disks, while both look mildly disturbed
in their outer part. The asymmetric H{\sc i} distribution of NGC~4568 with a maximum of H{\sc i} 
surface density to the west and a tail to the north are signs of an ongoing tidal interaction (Chung et al. 2009).
We detect radio continuum emission from both disks at 20 and 6~cm with a normal spectral index. Polarized emission at 20~cm is only detected 
along the minor axis of NGC~4567 east of the galaxy center. Surprisingly, this is the far side of the disk halo, because the
eastern side is the disk's near side. Disk-wide polarized emission at 6~cm is detected in both galaxies.
The magnetic field of the northern disk of NGC~4567 is that of a symmetric spiral galaxy.
Since the two galaxies are partially overlapping, the polarized emission of the southern disk of NGC~4567 is dominated by the emission
of NGC~4567, because the magnetic field in this region is the continuation of NGC~4567's magnetic field. 
The 6~cm polarized emission of NGC~4568 is highly asymmetric with its maxima south and west of the galaxy center.
The associated magnetic field is not azimuthal, but has a strong radial and/or vertical component. The magnetic field structure is comparable to 
that of NGC~4294 which forms a pair with NGC~4299 and that of the Antennae (NGC~4038/9; Chy\.{z}y \& Beck 2004).

\subsubsection{NGC~4579}

This giant Virgo Sa galaxy has a prominent bar and was classified as anemic by Cayatte et al. (1994).
It has a well-known Seyfert 2 nucleus, with radio jets (Contini 2004).
Star formation takes place only in a narrow ($20''$) ring structure at the end of the bar. Nevertheless, NGC~4579 has a high surface brightness
radio continuum disk which is symmetric. The spectral index in the disk region is quite shallow ($-0.6$). 
The disk is less extended to the north at 20~cm than at 6~cm.
Polarized emission at 20 and 6~cm is detected in the entire radio continuum disk. The degree of polarization at 6~cm is
one of the highest ($\sim 14$\,\%) in our sample. The ubiquitous polarized emission
at 20~cm is the exception in our sample. The magnetic field structure is very symmetric, a beautiful example of a regular field pattern
in a spiral galaxy with a strong bar, like NGC~1097 (Beck et al. 2005). However, the polarized emission is very smooth and not concentrated in
upstream or interarm regions. The smoothness could be an effect of the limited resolution of our observations.

\subsubsection{NGC~4689}

This small face-on spiral has a mildly truncated H{\sc i} disk to within the stellar disk, with a moderate H{\sc i} deficiency (Chung et al. 2009). 
NGC~4689 has a symmetric radio continuum disk at 20 and 6~cm with a low surface brightness ($\sim 200$~$\mu$Jy/beam at 6~cm).
The spectral index is shallower in starforming regions of the inner disk than in the outer disk. Polarized emission is marginally
detected in the galactic disk at 20 and 6~cm. The degree of polarization is intermediate ($\sim 10$\,\%).

\subsubsection{NGC~4713}

As NGC~4532, the face-on spiral galaxy NGC~4713 hosts a large H{\sc i} envelope (Chung et al. 2009). 
It also shares many H{\sc i} properties with NGC~4808, one of its nearest large neighbors in the southern outskirts of the cluster.
The H{\sc i} velocity field suggests a disturbed warp for the gas beyond the stellar disk (Chung et al. 2009).
Its radio continuum disk is symmetric at 20 and 6~cm with a
normal spectral index. Polarized emission at 20~cm is only detected in the southern part of the galactic disk.
Unlike NGC~4532, the polarized emission of NGC~4713 at 6~cm is also symmetric with a nearly perfect spiral structure of the
magnetic field. The strong magnetic arms are located mostly outside the main optical emission.
The degree of polarization is intermediate ($\sim 10$\,\%) for an unperturbed spiral galaxy.

\subsubsection{NGC~4808}

The third galaxy with an extended H{\sc i} envelope is NGC~4808. It shares many H{\sc i} properties with NGC~4713.
There is strong star formation throughout the stellar disk, with an asymmetric and patchy distribution.
Its radio continuum properties are comparable to those of NGC~4532.
This highly inclined galaxy has a symmetric radio continuum disk at 20 and 6~cm with a vertically extended radio halo.
Polarized emission at 20~cm is only detected in the eastern part of the galactic disk. The polarized emission distribution at 6~cm
is asymmetric with its maximum located southeast of the galaxy center.  The degree of polarization is low ($\sim 6$\,\%).
The magnetic field is almost everywhere perpendicular to the galactic disk.

\section{Galaxy classes based on the radio continuum properties \label{sec:galprops}}

Based on the radio continuum properties, we can divide our Virgo cluster sample into 5 classes:  
galaxies with (i) asymmetric polarized emission at 20~cm, (ii)  a symmetric magnetic field spiral pattern,
(iii) a radio continuum halo, (iv) an asymmetric and/or truncated radio halo, and (v) asymmetric 6~cm polarized emission.
Classes (ii) to (v) are based on our 6~cm data (Table~\ref{tab:classes}). Whereas the radio continuum morphologies of classes
(i) to (iii) are also observed in field galaxies, those of classes (iv) and (v) are caused by tidal interactions
and/or ram pressure in the cluster environment. In the following, we
describe these classes and put them into context. We also include the 8 Virgo spiral galaxies presented in Vollmer et al. (2010)
in our analysis.
\begin{table*}
      \caption{Galaxy classes based on radio continuum morphology.}
         \label{tab:classes}
      \[
         \begin{array}{lcl}
           \hline
           \noalign{\smallskip}
           {\rm Class} & {\rm wavelength} & {\rm galaxies\ (NGC})\\
       \noalign{\smallskip}
       \hline
       {\rm asymmetric\ polarized\ emission} & {\rm 20~cm} & 4302, 4303, 4321, 4419, 4579, 4689, 4713 \\
       \noalign{\smallskip}
       \hline
       {\rm symmetric\ magnetic\ field\ spiral\ pattern} & {\rm 6~cm} & 4303, 4579, 4713 \\
       \noalign{\smallskip}
       \hline
       {\rm radio\ continuum\ halo} & {\rm 6~cm} & 4178, 4192, 4294, 4302, 4330, 4419, 4532, 4808 \\
       \noalign{\smallskip}
       \hline
       {\rm asymmetric\ and/or\ truncated\ radio\ halo} & {\rm 6~cm} & 4330, 4419, 4424, 4457 \\
       \noalign{\smallskip}
       \hline
       {\rm asymmetric\ 6~cm\ polarized\ emission} & {\rm 6~cm} & 4294, 4298, 4457, 4532, 4568, 4808 \\
       \noalign{\smallskip}
       \hline
  \end{array}
	 \]
\end{table*}

\subsection{Asymmetric polarized emission at 20~cm \label{sec:asympol20}}

Braun et al. (2010) found a remarkable pattern in their Westerbork 20~cm sample relating to the basic distribution of polarized intensity in galaxy disks. 
There is a general gradient in the average polarized intensity that is approximately aligned with the major axis of the target 
galaxy. This gradient from high to low polarized intensity has the same sign in all well-detected cases, from high values on the kinematically approaching 
major axis to low values on the receding major axis. The polarized intensity thus shows a global minimum toward the receding major axis of the galactic disk.
Braun et al. (2010) stated that the effect cannot be explained by a symmetric planar field geometry.
Instead, five ingredients are needed: (i) an axissymmetric spiral structure (ASS) with not too small pitch angles ($>10^{\circ}$),
(ii) the spiral structure is trailing, (iii) a thick magnetic disk (several kpc) with a quadrupole topology (Fig.~10 of Braun
et al. 2010), (iv) polarized emission of the far side of the thick disk is not detected due to Faraday depolarization, and (v)
and moderate inclination angles (about $20^{\circ}$ to $75^{\circ}$). 
If these conditions are fulfilled, the polarized emission at 20~cm shows a minimum at the receding side and a maximum at the
approaching side of the galactic disk. To determine the approaching side, we used the H{\sc i} velocity fields of  Chung et al. (2009)
and Cayatte et al. (1990).
We have one galaxy in common with the Braun et al. sample, NGC~4321 (Fig.~1 of Vollmer et al. 2010), 
where this effect is present: polarized radio continuum emission is only observed in the northeastern half of the disk which is 
the approaching side.

\begin{figure*}
  \centering
  \resizebox{15cm}{!}{\includegraphics{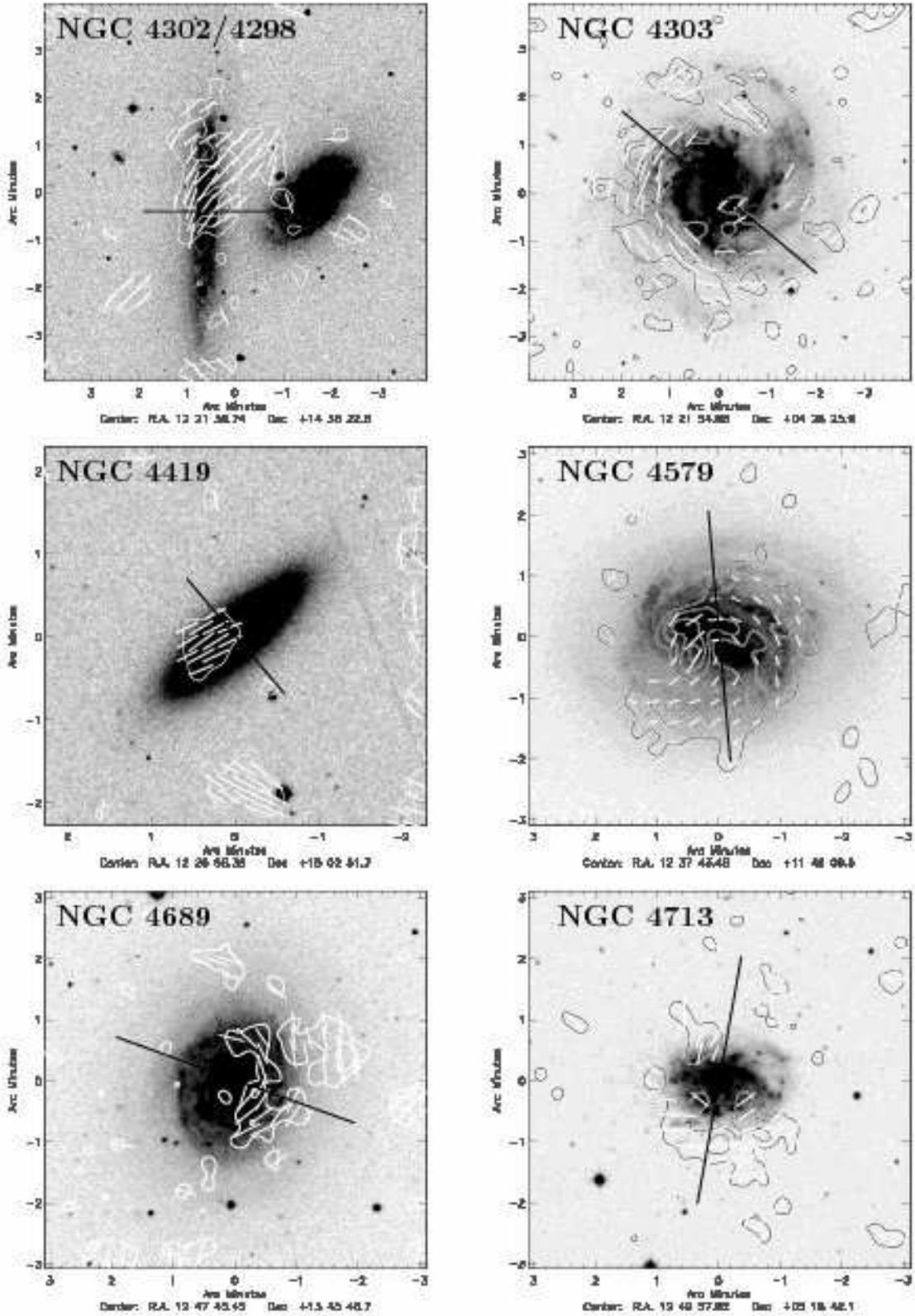}}
  \caption{Virgo cluster spiral galaxies with 20~cm polarized emission mainly found at the approaching side of the galactic disk.
    For all galaxies, except NGC~4302 which is seen edge-on, this asymmetry is probably
    caused by a quadrupole ASS field topology of the large-scale magnetic field (Braun et al. 2010).
    Greyscale: optical, contours: 20~cm polarized radio continuum, lines: apparent B field vectors
    whose sizes are proportional to the polarized intensity.
    The contour levels are $(3,5,8,12,20,30,50,80,120,200,300)$ times the rms noise levels
    from Table~\ref{tab:table}. The black line indicate the galaxy's minor axis.
  \label{fig:zusammen1braun}}
\end{figure*}

In our new sample we found 5 cases (Fig.~\ref{fig:zusammen1braun}): (1) the polarized emission distribution is
strongest in the approaching southwestern disk of the nearly face-on ($i=25^{\circ}$) galaxy NGC~4303, (2) polarized emission is only detected in the 
approaching southwestern part of the highly inclined ($i=74^{\circ}$) disk of NGC~4419, (3) the more face-on ($i=38^{\circ}$) spiral galaxy NGC~4579 
displays polarized emission mainly in the approaching western part of the galactic disk, (4) the polarized emission of the more face-on ($i=37^{\circ}$) 
spiral galaxy NGC~4689 is mainly detected in the northwestern part of the disk which is approaching, (5) in NGC~4713 ($i=52^{\circ}$) the northeastern side 
of the galactic disk is approaching where 20~cm polarized emission is detected, in addition polarized
emission is also found in the south of the galaxy center which is receding. We thus approximately double the existing sample of
spiral galaxies with asymmetric 20~cm polarized emission distribution showing a global minimum at the receding side of the galactic disk and
hence strengthening the evidence of dominating quadrupolar-type magnetic fields in spiral galaxies.
In addition, the highly inclined ($i=90^{\circ}$) spiral galaxy NGC~4302 also displays 20~cm polarized emission only in the northern approaching part of the disk.
Because of its high inclination, this behavior must be due to different unknown mechanism.

\subsection{Symmetric magnetic field spiral pattern}

\begin{figure*}
  \centering
  \resizebox{\hsize}{!}{\includegraphics{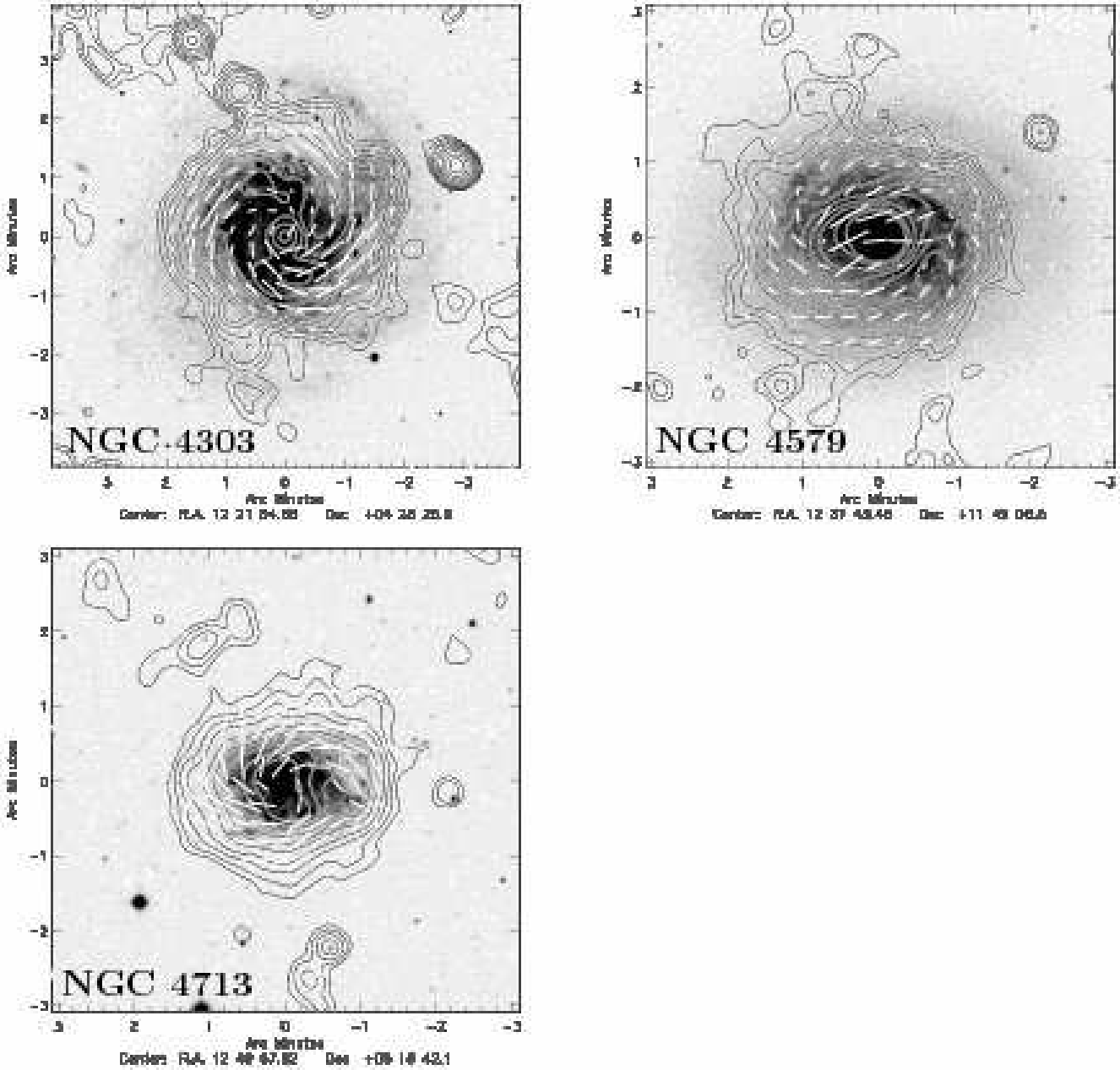}}
  \caption{Virgo cluster galaxies with symmetric radio continuum distributions and
    a symmetric spiral pattern of the apparent magnetic field vectors.
    Greyscale: optical, contours: 6~cm radio continuum, lines: apparent B field vectors
    whose sizes are proportional to the polarized intensity.
    The contour levels are $(3,5,8,12,20,30,50,80,120,200,300)$ times the rms noise levels
    from Table~\ref{tab:table}.
  \label{fig:zusammen1spirals}}
\end{figure*}

Symmetric magnetic field spiral patterns, as frequently observed in field spiral galaxies, are the exception in the Virgo cluster.
Only four out of 27 observed Virgo spiral galaxies have symmetric spiral patterns: NGC~4303, NGC~4579, NGC~4713 (Fig.~\ref{fig:zusammen1spirals})
and NGC~4321 (Vollmer et al. 2010). The projected distances from the cluster center range from $1.8^{\circ}$(0.5~Mpc) to $8.5^{\circ}$(2.5~Mpc).
NGC~4321 and NGC~4303 resemble healthy field spirals. Whereas NGC~4713 has an extended H{\sc i}
envelope, i.e. it has an excess of atomic hydrogen compared to field spirals, NGC~4579 is an anemic spiral galaxy with an H{\sc i}
deficiency of $0.95$ (Chung et al. 2009). The degree of polarization and the $6$~cm surface brightness are high in these galaxies
($> 10$\,\% and $>300$~$\mu$Jy/beam). The symmetric structure of the magnetic field in NGC~4713 indicates that the outer H{\sc i}
envelope has no disturbing influence on the inner disk. The outer H{\sc i} disk might thus be qualified as settled.
It is surprising that an anemic, highly H{\sc i} deficient galaxy as NGC~4579 shows radio
continuum properties of a normal field spiral galaxy. We come back to this point in Sect.~\ref{sec:radiofir}.

\subsection{Radio continuum halos}

\begin{figure*}
  \centering
  \resizebox{12cm}{!}{\includegraphics{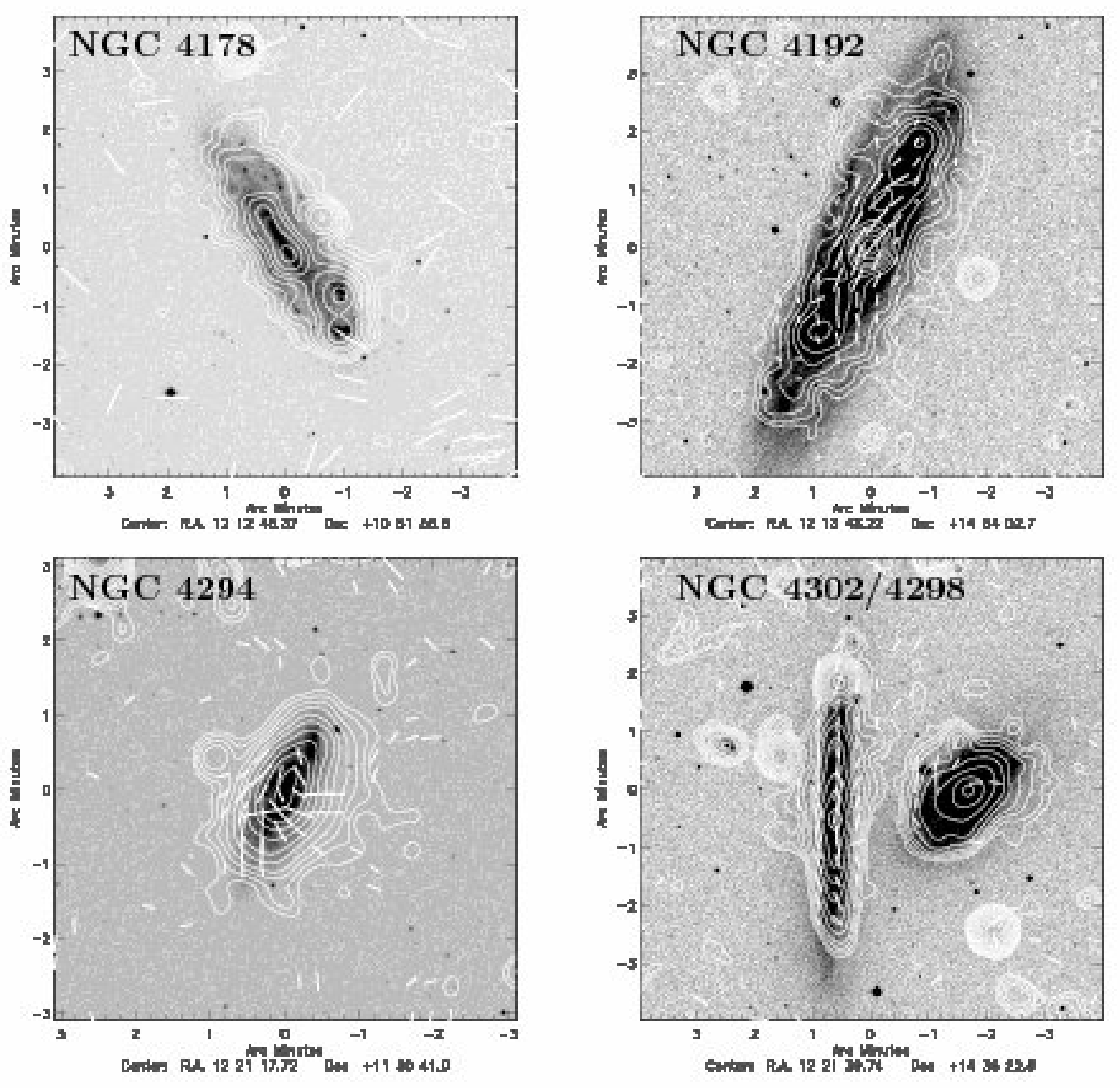}}
  \resizebox{12cm}{!}{\includegraphics{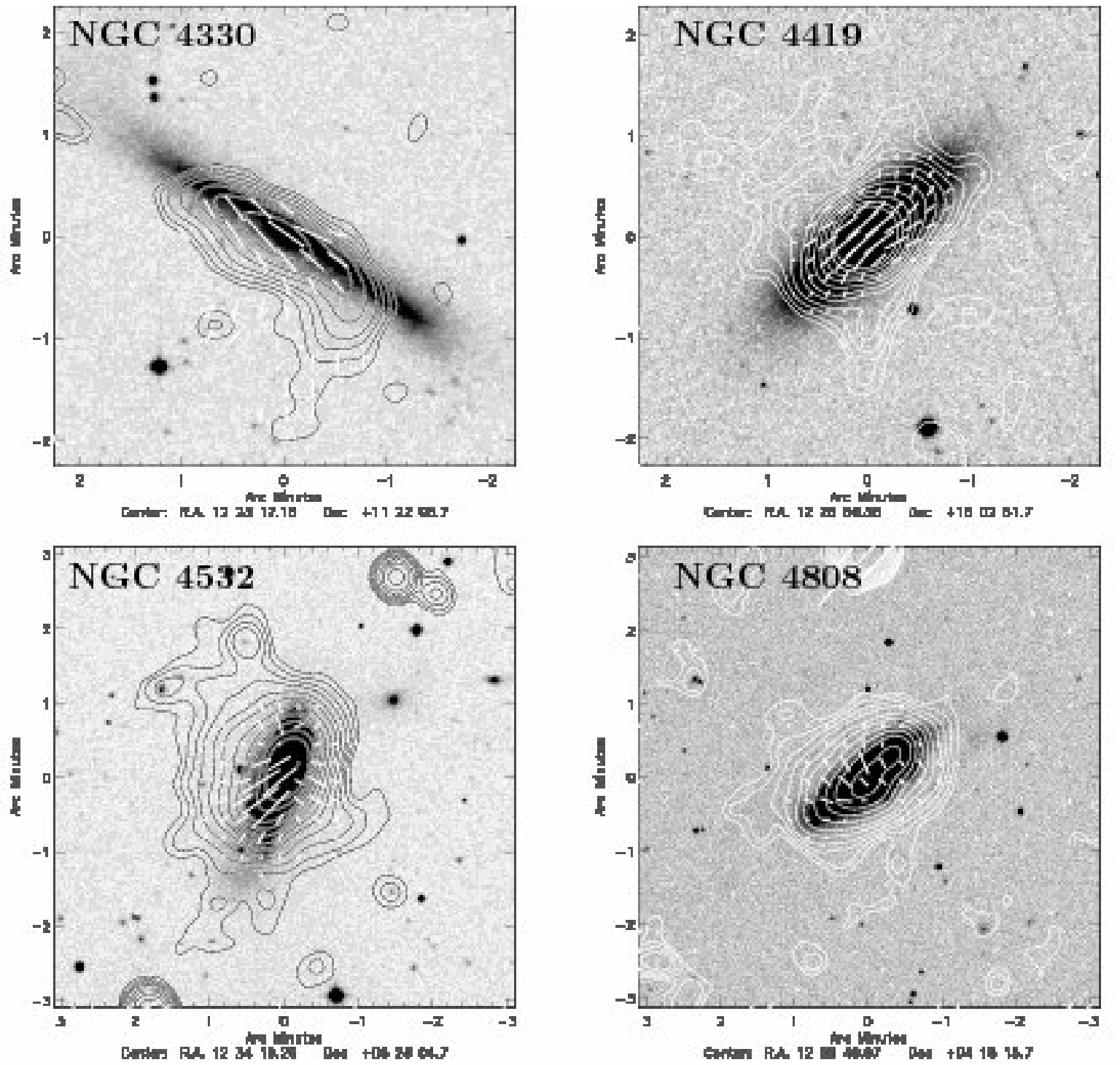}}
  \caption{Highly inclined Virgo cluster galaxies with extended symmetric radio halos.
    Greyscale: optical, contours: 6~cm radio continuum, lines: apparent B field vectors
    whose sizes are proportional to the polarized intensity.
    The contour levels are $(3,5,8,12,20,30,50,80,120,200,300)$ times the rms noise levels
    from Table~\ref{tab:table}.
  \label{fig:zusammen1halos1}}
\end{figure*}

Vertically extended radio continuum halos are observed in all highly inclined ($i \ge 70^{\circ}$) Virgo spiral galaxies
(Fig.~\ref{fig:zusammen1halos1}), except in NGC~4216 where the radio continuum disk is thin.
NGC~4396, NGC~4402, and NGC~4438 are also members of this class (Vollmer et al. 2010).
Radio continuum halos are one component of a multiphase interstellar medium which is expelled by the common
action of multiple supernova explosions. The cosmic ray electrons and magnetic fields are lifted to higher latitudes together with
neutral (H{\sc i}, e.g., Oosterloo et al. 2007) and ionized (H$\alpha$, e.g., Rossa \& Dettmar 2003) material.
The existence and size of a radio continuum halo depends on the local star formation activity per unit surface and
the gravitational potential which has to be overcome to launch a galactic wind (Dahlem et al. 2006).
We note that the galaxy with the lowest $6$~cm surface brightness, NGC~4216, indeed does not have a radio halo.
A detailed investigation of the halo properties and their relation to the properties of the galactic disks is
beyond the scope of this article and will be presented in a subsequent publication.
The projected magnetic field vectors of most of the symmetric halos show a characteristic X structure (NGC~4192,
NGC~4302, NGC~4419, NGC~4532). In addition, the magnetic field of NGC~4808 is dominated by the vertical component,
which can be seen as an extreme X structure. 

\subsection{Asymmetric and/or truncated radio halos}

\begin{figure*}
  \centering
  \resizebox{\hsize}{!}{\includegraphics{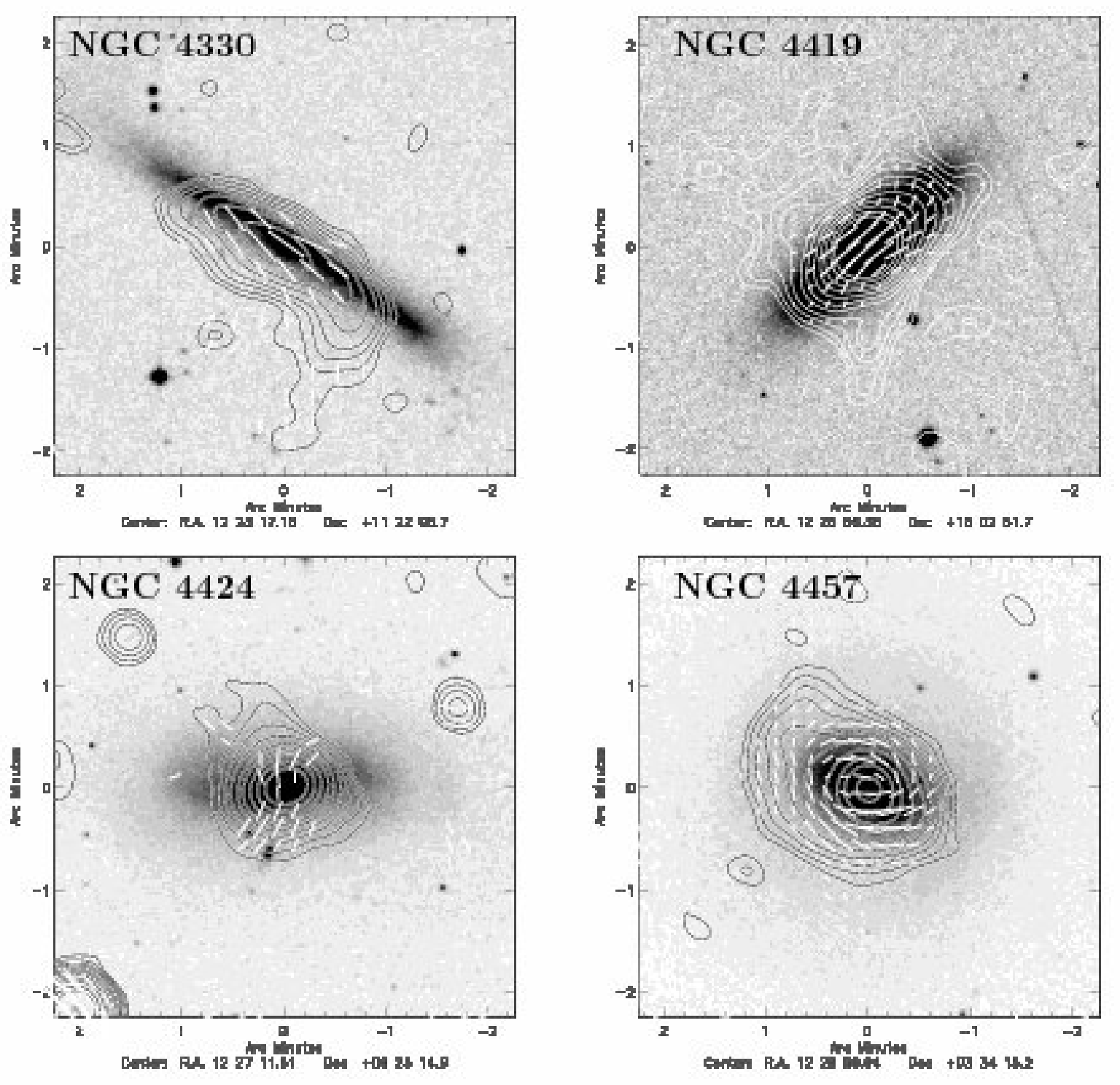}}
  \caption{Highly inclined Virgo cluster galaxies with asymmetric and/or radially truncated radio halos.
    Greyscale: optical, contours: 6~cm radio continuum, lines: apparent B field vectors
    whose sizes are proportional to the polarized intensity.
    The contour levels are $(3,5,8,12,20,30,50,80,120,200,300)$ times the rms noise levels
    from Table~\ref{tab:table}.
  \label{fig:zusammen1asymm}}
\end{figure*}

Five galaxies of our sample of 27 Virgo galaxies show an asymmetric radio continuum halo:
NGC~4330, NGC~4457 (Fig.~\ref{fig:zusammen1asymm}) and NGC~4396, NGC~4402, NGC~4438 (Vollmer et al. 2010).
Except NGC~4457, all galaxies are located in projection within the inner $3^{\circ}$(0.9~Mpc) of the cluster.
NGC~4438 is exceptional, because its one-sided extraplanar radio continuum emission is round and has a high
surface brightness ($\sim 1$~mJy/beam). It is thus not a classical radio halo, but most probably part of the
ram-pressure stripped interstellar medium of NGC~4438 (Vollmer et al. 2009).
NGC~4402 and NGC~4330 both have compressed radio halos on one side of the galactic disk. The cosmic
ray electrons and magnetic fields on the opposite side are probably pushed by ram pressure to higher latitudes
(Crowl et al. 2005, Vollmer et al. 2012a). NGC~4396 shows a radio continuum and H{\sc i} tail to the northwest.
This gas tail is probably also due to ram pressure (Chung et al. 2007).

The strongly truncated radio halos of NGC~4419 and NGC~4424 come along with strongly truncated H{\sc i} disks
(Chung et al. 2009). In these galaxies the gas and star formation activity are radially truncated.
Therefore, an untruncated gas and starforming disk is not a requisite for the development of a vertically extended
radio continuum halo. In NGC~4424, the radio halo stems from a nuclear outflow powered by bright central H{\sc ii}
regions. Since the southern H{\sc i} tail is most probably due to ongoing ram pressure stripping (Chung et al. 2007),
the northern outflow resists ram pressure, i.e. it is sufficiently overpressured. Another example
of an overpressure outflow in a Virgo cluster spiral galaxy affected by ram pressure is NGC~4569 (Chy\.{z}y et al. 2006).

The only azimuthally asymmetric radio continuum total power distribution is found in NGC~4457.
A radio continuum arm without an optical nor H{\sc i} counterpart is observed to the northeast within the optical radius 
$R_{25}$. The magnetic field pattern is symmetric in the inner disk and ordered in the continuum arm.
The H$\alpha$ emission shows a one-arm structure in the inner disk located south of the
galaxy center. The only viable interpretation of these features is a minor merger. This is surprising, because
a small satellite has accreted onto a strongly H{\sc i}-deficient galaxy with a strongly truncated gas disk.
The merger led to a vigorously starforming spiral arm from which cosmic ray electrons and the associated
magnetic field were expelled vertically and radially. The radial expulsion is only possible because of the
truncation of NGC~4457's gas disk. The azimuthal offset between the H$\alpha$ and radio continuum arm can be explained by differential
rotation and a strongly varying star formation rate in the spiral arm.

\subsection{Asymmetric 6~cm polarized emission \label{sec:asympol}}

\begin{figure*}
  \centering
  \resizebox{15cm}{!}{\includegraphics{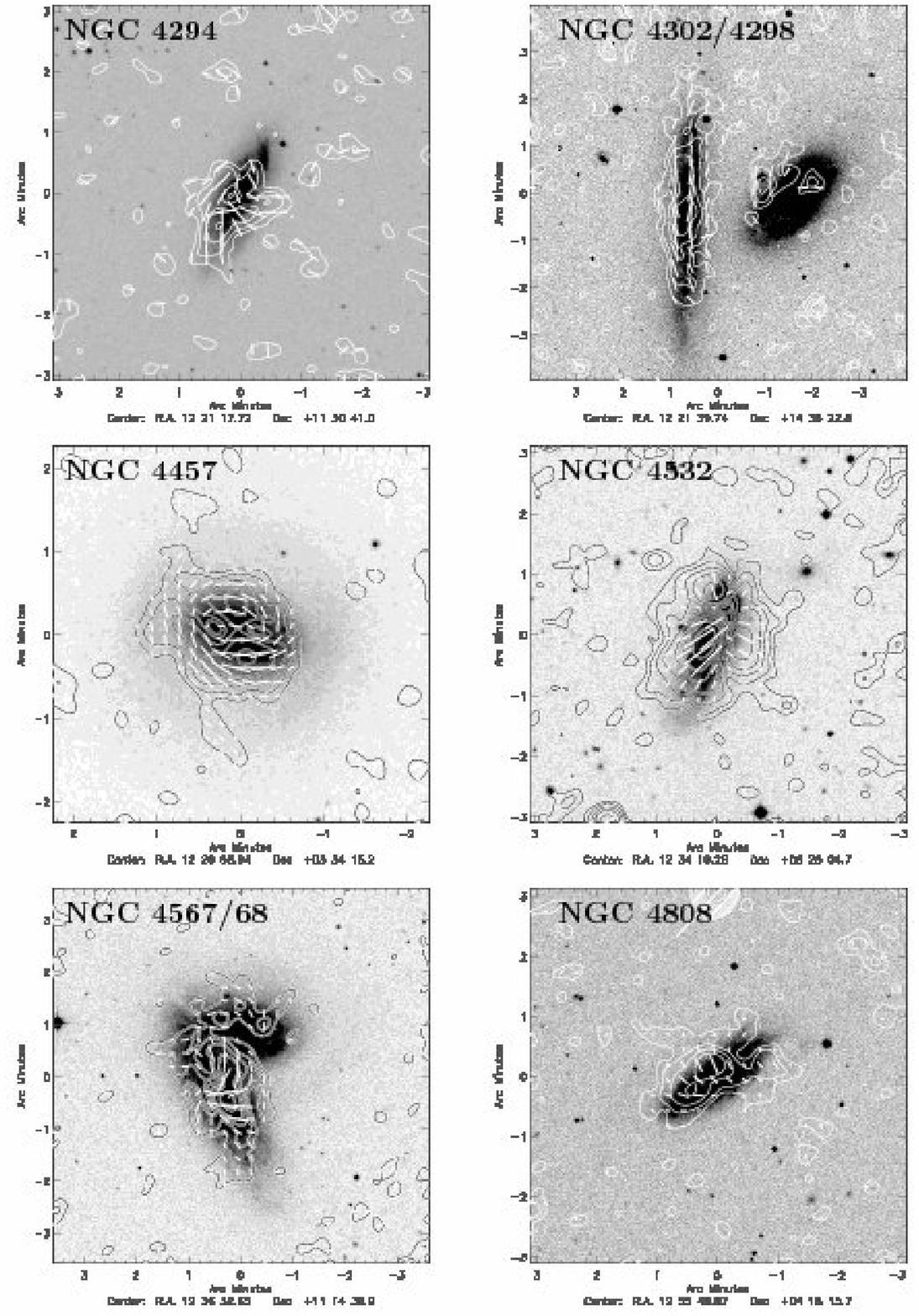}}
  \caption{Virgo cluster galaxies with asymmetric distributions of the 6~cm polarized emission.
    Greyscale: optical, contours: 6~cm polarized radio continuum, lines: apparent B field vectors
    whose sizes are proportional to the polarized intensity.
    The contour levels are $(3,5,8,12,20,30,50,80,120,200,300)$ times the rms noise levels
    from Table~\ref{tab:table}.
  \label{fig:zusammen1pionopt}}
\end{figure*}

Asymmetric 6~cm polarized radio continuum distributions are observed in 6 out of 19 galaxies of the new sample
(NGC~4294, NGC~4298, NGC~4457, NGC~4532, NGC~4568, NGC~4808; Fig.~\ref{fig:zusammen1pionopt}).
They are located at various projected distances from the cluster center.
Vollmer et al. (2007) claimed that asymmetric ridges of polarized radio continuum emission located in the
outer parts of the galactic H{\sc i} disks in Virgo cluster spiral galaxies are signs of ongoing ram-pressure compression
(NGC~4388, NGC~4402, NGC~4438, NGC~4501, NGC~4522, NGC~4654).
In addition, these polarized ridges are located at the outer edge of H{\sc i} disks.
Is this also the case for the 6 galaxies of the new sample? 

The asymmetry of the polarized radio continuum emission distribution of NGC~4457 is due to the northeastern arm.
The distribution thus does not resemble those of spiral galaxies only affected by ram pressure (NGC~4388, NGC~4402, 
NGC~4501, NGC~4522).
NGC~4294 shares a common H{\sc i} envelope with NGC~4299. However, the stellar distributions of the two galaxies
are symmetric. The observed asymmetry in polarized radio continuum emission in NGC~4294 is located well inside
the H{\sc i} distribution. We interpret this asymmetry as trace of a tidal interaction between the two galaxies.
In contrast to the previous cases, the maxima of polarized emission are located at the edge of the H{\sc i} disk only in 
NGC~4298 (Fig.~\ref{fig:zusammen1n4298}) and NGC~4568 (Fig.~\ref{fig:zusammen1n4567}). 
NGC~4568 is part of a tidally interacting pair. NGC~4298 is also affected by the tidal field of NGC~4302 (Vollmer et al. in prep.). 
Therefore, the asymmetries in the polarized radio continuum emission of these two galaxies are likely due to tidal compression and/or shear.

NGC~4532 and NGC~4808 possess huge disturbed H{\sc i} envelopes with complex dynamics (Chung et al. 2009). We therefore 
interpret the observed asymmetry in polarized radio continuum emission as the result of external gas accretion which
gives rise to shear and compressive gas motions.
We thus conclude that none of the Virgo galaxies of the new sample, except NGC~4330, is affected by strong 
ongoing ram pressure stripping.
The sample of Vollmer et al. (2007) had thus been selected with great care. The vast majority of cases of strong ram pressure stripping
were already contained in this sample. However, the case of NGC~4330 shows
that even if ram pressure stripping is ongoing, the ridge of polarized emission cannot be observed if compression occurs
in the plane of the sky. This has to be check with numerical simulations including ram pressure and magnetic fields (e.g., Vollmer et al. 2012a).
In addition, the simultaneous presence of strong shear motions can erase the signature of mild compression on the polarized
radio continuum emission (NGC~4254;, Chy\.zy 2008; NGC~4569, Chy\.zy priv. comm.). Thus, the presence of an asymmetric ridge of polarized radio continuum emission 
located in the outer parts of truncated H{\sc i} disks in Virgo cluster spiral galaxies is a signpost of ongoing ram pressure stripping.
On the other hand, the absence of such a polarized radio ridge might be due to unfavorable projection or strong shear motions of the gas
in the galactic disk. An alternative way to detect the presence of ongoing ram pressure stripping that is independent of projection and
shear motions is to search for regions of low radio/FIR ratio (compared to the mean) at the compressed side of the disk 
(NGC~4330, NGC~4402, NGC~4522, NGC~4569; Murphy et al. 2009). 

Pfrommer \& Dursi (2010) interpreted the polarized radio ridges as the draping of the cluster magnetic field around
the galactic disk. The draping model predicts no relation between the disk orientation and the observed magnetic field orientation,
because the cluster magnetic field is independent of disk orientation.
Since in our sample of 28 Virgo spiral galaxies the magnetic fields within the polarized 
radio ridges are all azimuthal and that of NGC~4501 is clearly linked to galactic structure (Vollmer et al. 2008), we do not
see any evidence for magnetic draping in our sample galaxies. If the orientation of the magnetic
fields in the polarized radio ridges is caused by ram pressure compression, they are approximately perpendicular to the
motion of the galaxy within the cluster. 
Pfrommer \& Dursi's interpretation of the direction of the observed field lines as a dominating radial magnetic cluster field
is not supported by the data. Instead, the locations of the polarized ridges within the galactic disks are determined by 
the eccentric orbits of the spiral galaxies within the Virgo cluster (Dressler 1986).
Moreover, since there is no connection between the ISM magnetic field and the large-scale draped cluster magnetic field,
cosmic electrons which are formed inside the ISM magnetic field cannot penetrate into the cluster magnetic field to illuminate it.
The only way to obtain radio continuum emission involving the draped cluster magnetic field is that 
supernovae have to explode within the draped cluster magnetic field  to provide the cosmic-ray electrons. 
This is only possible if ram pressure pushes the ISM fast enough to smaller galactic radii, so that massive stars which were
formed in the ISM compression region subsequently explode as supernovae in a gas-free region. The action of ram pressure
is thus needed to make a putative draped cluster magnetic fields visible in synchrotron emission. 
Even if magnetic draping occurs as predicted by Pfrommer \& Dursi (2010), we expect that ISM compression by ram pressure
is the dominant mechanism producing the observed asymmetries in polarized radio continuum emission.

A second anomaly of the majority of galaxies with asymmetric 6~cm polarized radio continuum emission are dominating
vertical magnetic field components in four out of six galaxies (NGC~4294, NGC~4532, NGC~4568, NGC~4808; Fig.~\ref{fig:zusammen1pionopt}).
The non-detection of plane-parallel regular magnetic fields from the thin disk might be due to an absence of these large-scale fields or
Faraday depolarization even at 6~cm. We note that all four galaxies are interacting tidally (NGC~4294 and NGC~4568) or with an
accreting gas halo (NGC~4532 and NGC~4808). Both effects can be achieved by enhanced small-scale turbulence.
We thus conclude that the interaction leads to the destruction of the regular magnetic field and/or strong Faraday depolarization most
probably caused by strong small-scale gas turbulence.

\section{Integrated galaxy properties \label{sec:discussion}}

We saw that the observed galaxies have very different degrees of polarization\footnote{We do not take into account
the polarized emission at 20~cm, because it suffers severe Faraday depolarization} at 6~cm implying that the
dynamos responsible for the large-scale magnetic fields have very different efficiencies. 
Following Ruzmaikin et al. (1988) the galactic mean-field dynamo in disk galaxies allows the generation of a regular galactic
magnetic field as a result of joint action of differential rotation $\Omega$ and helical turbulent motions of interstellar gas
(see also Arshakian et al. 2009). The latter is responsible for the so-called $\alpha$-effect.
The joint action of both generators can be described by a so-called dynamo number:
\begin{displaymath}
\vert D_{\rm d}\vert \simeq 9 \left({{H \Omega} \over v}\right)^2 \ ,
\label{eq:dynamo}
\end{displaymath}
where $\Omega$ is the angular velocity, $H$ the gas disk height, and $v$ the gas turbulent velocity.
The regular field strength increases faster for higher dynamo numbers in the exponentially growing phase of a large-scale dynamo. 
The final field strength also depends on the dynamo saturation, which is a matter of heavy debate (e.g., Moss \& Sokoloff 2011).
The disk height and turbulent velocity are not independent. High turbulent velocities generally lead to
a thicker gas disk. In addition, the dynamo needs a minimum star formation activity to produce helical turbulent motions.
On the other hand, star formation should not be too active, because a high star formation activity leads to
high turbulent velocities making the dynamo inefficient.

To investigate the origin of the different dynamo efficiencies, we compare the following integrated galaxy properties: 
\begin{itemize}
\item
average resolved degree of polarization: integrated polarized flux divided by the total power flux.
This is different from the average degree of polarization based on single dish measurements which
only partly resolve the galaxies (Stil et al. 2009),
\item
mean total power surface brightness within the radio contours,
\item
mean spectral index based on the total power fluxes at 6~cm and 20~cm,
\item
FIR $70$~$\mu$m flux divided by the 6~cm total power flux.
\end{itemize}
We separate the 27 galaxies in the full sample into different groups:
\begin{itemize}
\item
galaxies affected by ram pressure,
\item
galaxies undergoing a tidal interaction,
\item
galaxies experiencing ram pressure and a tidal interaction,
\item
galaxies possessing a large accreting gas envelopes,
\item
galaxies not affected by an interaction.
\end{itemize}

\subsection{Average degree of polarization}

\begin{figure*}
  \centering
  \resizebox{10cm}{!}{\includegraphics{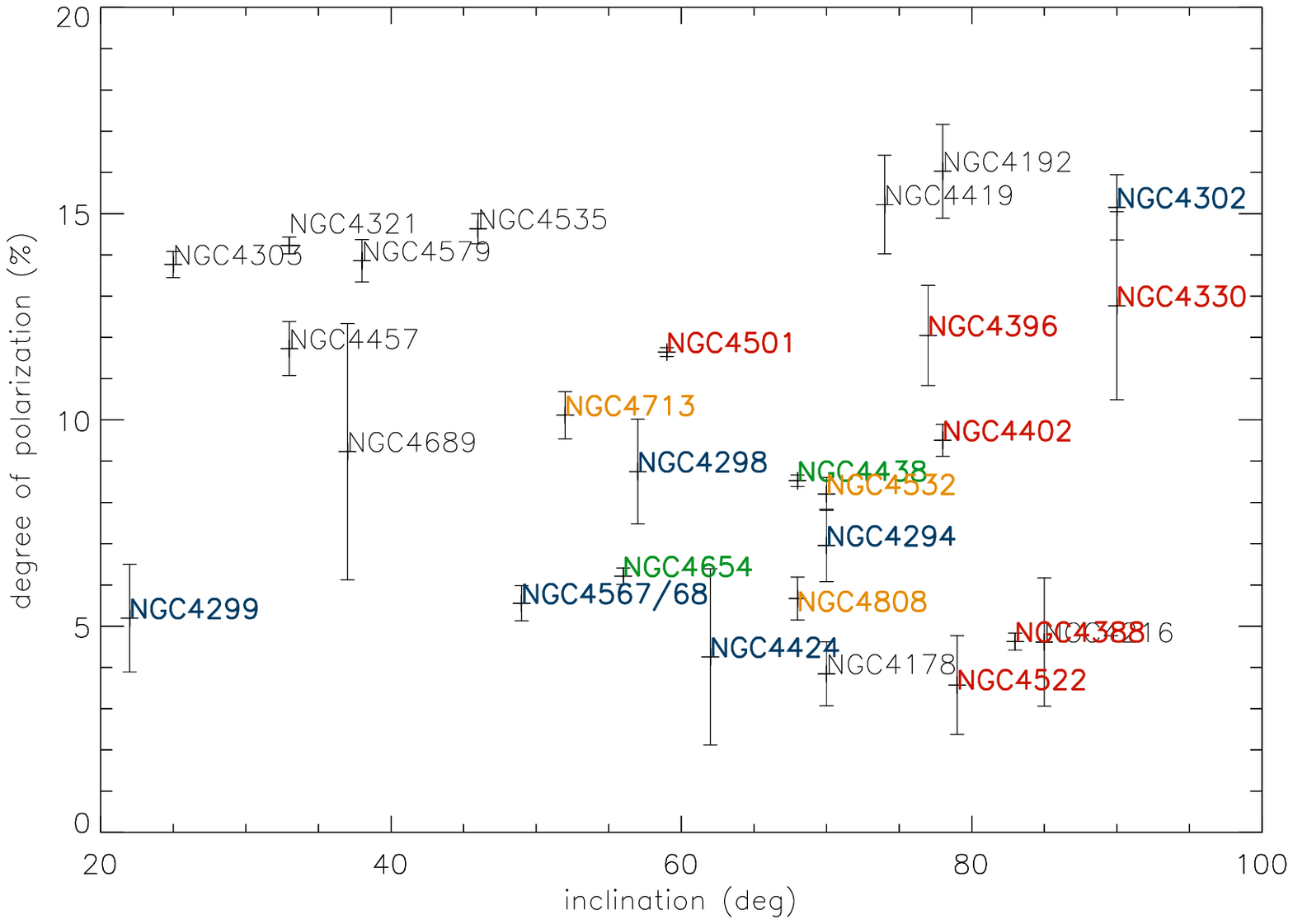}}
  \resizebox{10cm}{!}{\includegraphics{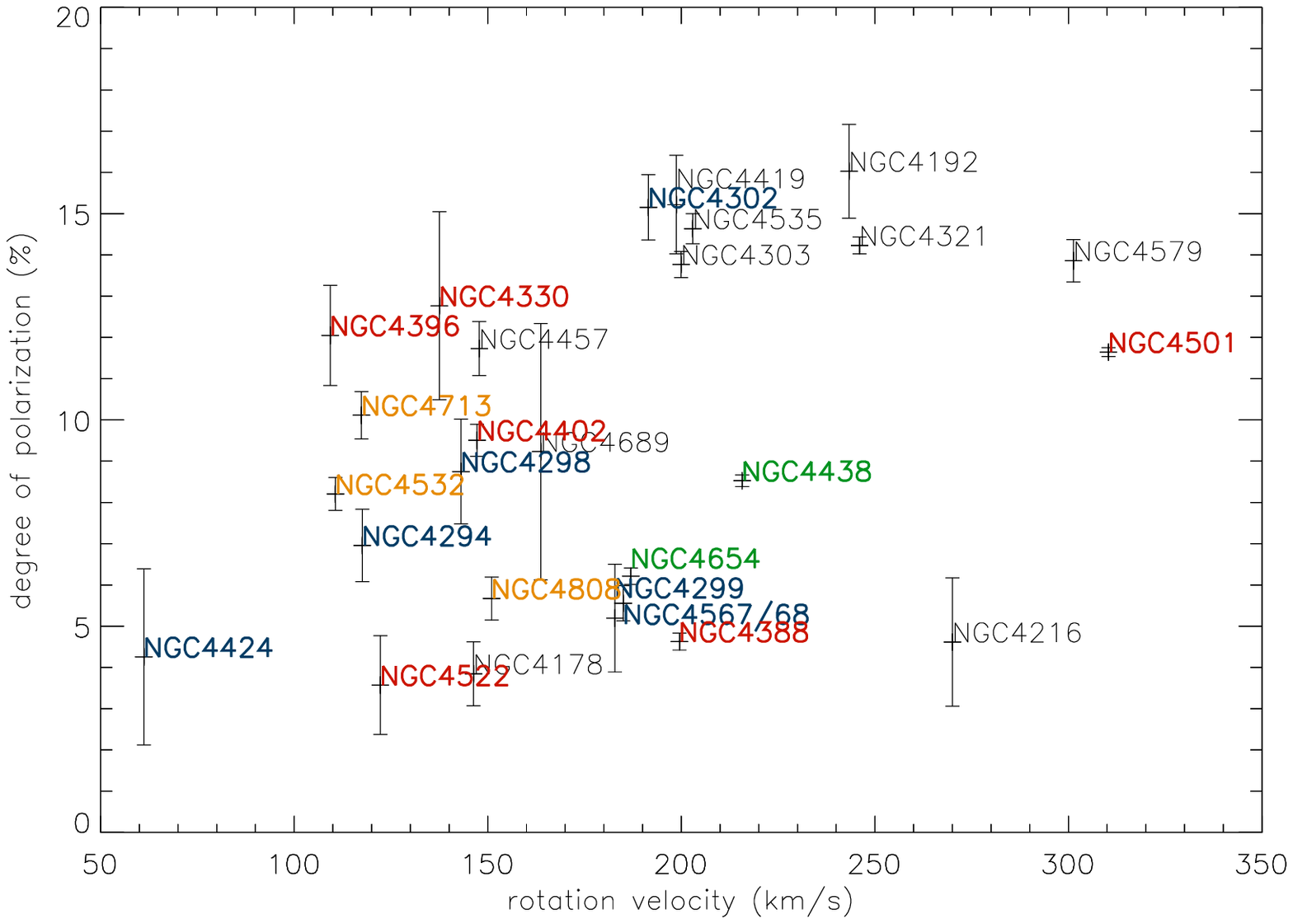}}
  \resizebox{10cm}{!}{\includegraphics{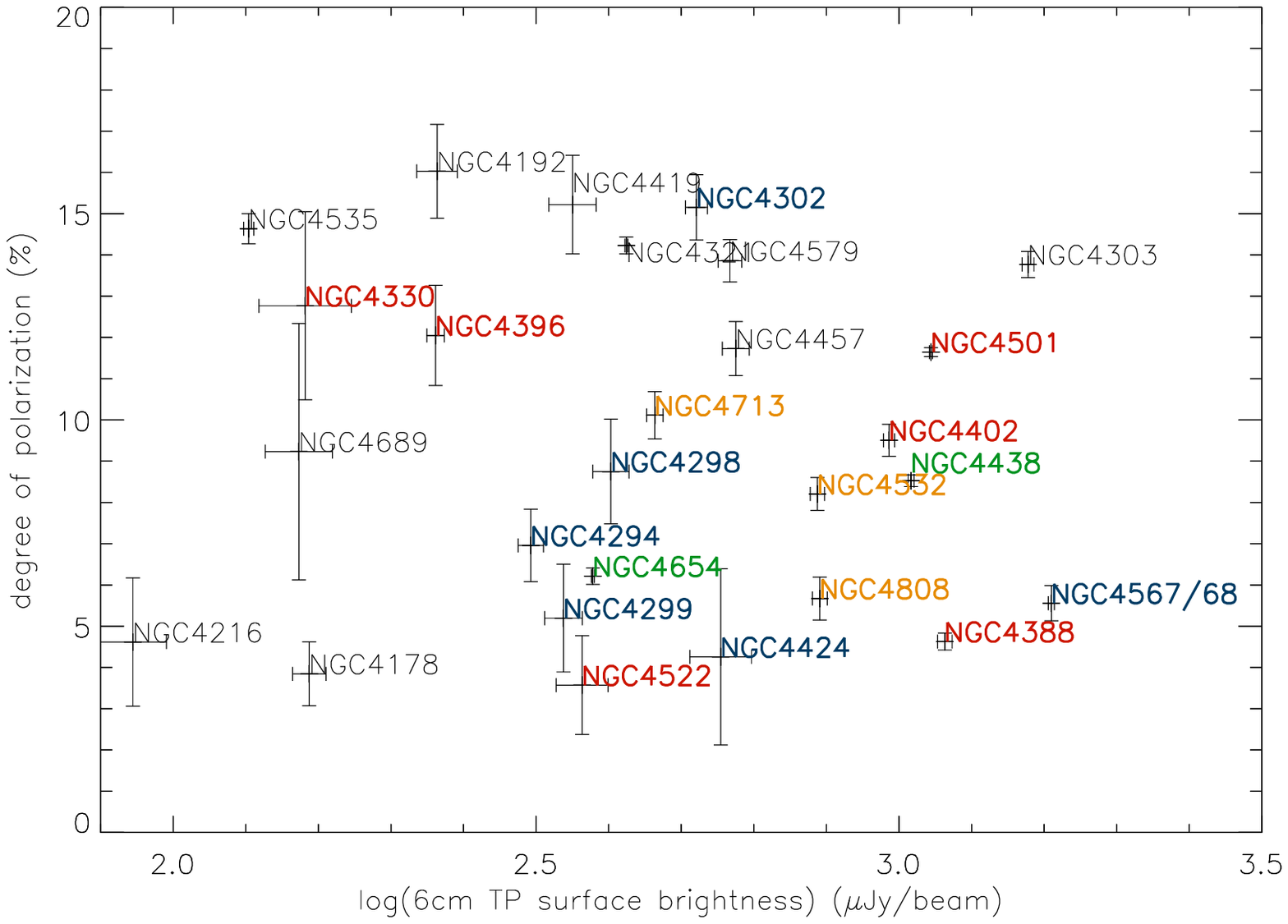}}
  \caption{Average degree of polarization at 6~cm as a function of galaxy properties. Upper panel: 
    inclination, middle panel: rotation velocity, lower panel: mean 6cm total power surface brightness
    within the radio isophotes. Red: galaxies experiencing ram pressure stripping, blue: galaxies
    experiencing a tidal interaction, green: galaxies experiencing a tidal interaction and ram pressure
    stripping, orange: galaxies with H{\sc i} envelopes, black: non-interacting.
  \label{fig:stats1}}
\end{figure*}

Since the polarized radio continuum emission of highly inclined spiral galaxies can suffer Faraday
depolarization with increasing distance along the major axis, we first look for a dependency of the
average degree of polarization on the disk inclination angle (upper panel of Fig.~\ref{fig:stats1}).
A very weak correlation is found between the inclination angle and the average resolved degree of polarization
with a Spearman rank coefficient of $-0.10$ and a significance value of $0.63$ (small value indicates a significant correlation).

As expected by dynamo theory, the average degree of polarization increases with increasing rotation velocity
(middle panel of Fig.~\ref{fig:stats1}). The Spearman rank coefficient is the highest ($0.37$, NGC~4216 removed: $0.47$) of all tested
correlation with the highest significance (significance value = $0.06$, NGC~4216 removed: $0.02$).
NGC~4216 is a clear exception of this correlation with a high rotation velocity 
($270$~km\,s$^{-1}$) and a small degree of polarization ($\sim 4.5$\,\%). This galaxy has a very low gas surface 
density and star formation activity. The latter property might be one reason for the low average degree of polarization.
Another reason might be a small ratio between the height of the gas disk and the turbulent velocity ($H/v$, Eq.~\ref{eq:dynamo}).
In NGC~4216, the gravitational potential of the galactic disk is dominated by the stars.
The vertical pressure equilibrium can thus be described as
\begin{equation}
\rho v^2 = \pi G \Sigma_{*} \Sigma v/v_{*}\ ,
\end{equation}
where $\rho$ is the gas density, $G$ the gravitation constant, $\Sigma_{*}$ the stellar surface density, 
$\Sigma=2 \rho H$ the gas surface density, and $v_{*}$ the stellar turbulent velocity. This leads to a
disk height 
\begin{equation}
\frac{H}{v}=\frac{v_{*}}{2\,\pi\,G\,\Sigma_{*}}\ .
\end{equation}
A small dynamo number is thus achieved with a high stellar surface density $\Sigma_{*}$ and/or a low
stellar velocity dispersion.

There is a whole bunch of galaxies at $v_{\rm rot}=200$~km\,s$^{-1}$ showing somewhat lower degrees of polarization
than expected: NGC~4299, NGC~4388, NGC~4567/68, and NGC~4654. One of them is experiencing ram pressure
(NGC~4388 with a strongly truncated gas disk) and the others are tidally interacting galaxies.
A lower degree of polarization was also found in the Antennae (Chy\.zy \& Beck 2004) and other 
interacting and merging systems (Drzazga et al. 2011).

To further investigate the role of star formation, which is approximately proportional to the mean 6~cm
surface brightness within the radio contours, we present the average degree of polarization as a function of the
mean total power surface brightness in the lower panel of Fig.~\ref{fig:stats1}.

Stil et al. (2009) found a lower degree of polarization for more radio-luminous galaxies.
Since galaxies radio-luminous galaxies generally have a high mean surface brightness, we might expect a decreasing average degree of 
polarization at high mean  surface densities. However, we do not find this trend in our data
(Spearman rank coefficient of $-0.07$).
The non-interacting galaxies of our sample, except NGC~4535, show an increasing average degree of
polarization with increasing mean surface brightness between $100$ and $300$~$\mu$Jy/beam and an
approximately constant degree of polarization for higher mean surface brightnesses. NGC~4535 shows a much higher average degree of polarization
with respect to its very low mean surface brightness, because of large-scale shear motions (Chung et al. 2009) that amplify the
magnetic field. We note that two galaxies with the smallest
average degree of polarization have the lowest mean surface brightnesses (NGC~4178 and NGC~4216).
At mean surface densities exceeding $300$~$\mu$Jy/beam the average degree of polarization stays approximately constant.
If the interacting galaxies are also taken into account the trend disappears, because many of
these galaxies have a lower average degree of polarization than expected from their mean surface brightness.

In the upper panel of Fig.~\ref{fig:stats2} we present the average degree of polarization as a function of the projected
cluster radius. As expected, the galaxies experiencing a tidal interaction and/or ram pressure are located within a
radius of $4^{\circ}=1.2$~Mpc around the cluster center, whereas the galaxies with accreting gas envelopes are
located at larger projected distances. There is no correlation between the average degree of polarization and the projected
cluster radius (Spearman rank coefficient of $0.09$). 
We identify a potential lack of galaxies with low average degrees of polarization ($<5$\,\%) at large
projected cluster radii ($>5^{\circ}$). In addition, the majority (4 out of 6) of the galaxies experiencing ram pressure have high
average degrees of polarizations ($\ge 9$\,\%), whereas most (5 out of 6) of the tidally interacting galaxies show low average
degrees of polarization  ($<9$\,\%).

We thus conclude that ram pressure stripping can decrease whereas tidal interactions most frequently decreases the average degree of 
polarization of Virgo cluster spiral galaxies. The low average degree of polarization in tidally interacting galaxies is
consistent with the lack of polarized emission from the thin disk in NGC~4294, NGC~4567/68 (Sect.~\ref{sec:asympol}).
The exception of this trend is NGC~4302 which interacts tidally with NGC~4298. A possible reason might be that the
galaxy orbit is almost perpendicular to the disk plane of NGC~4302 making tidal effects less efficient (Vollmer et al., in prep.).

\subsection{Spectral index and radio/FIR correlation \label{sec:radiofir}}

\begin{figure*}
  \centering
  \resizebox{10cm}{!}{\includegraphics{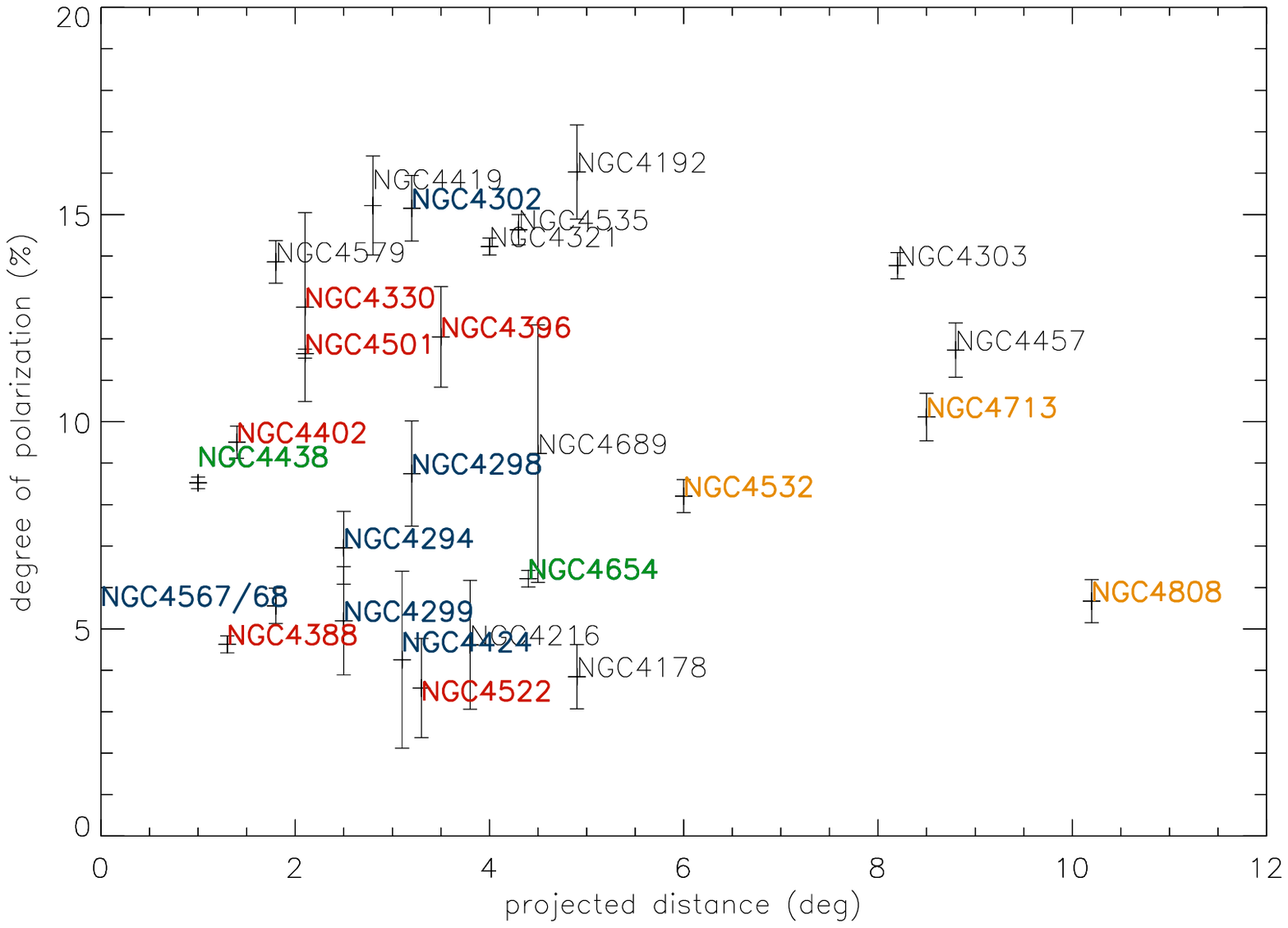}}
  \resizebox{10cm}{!}{\includegraphics{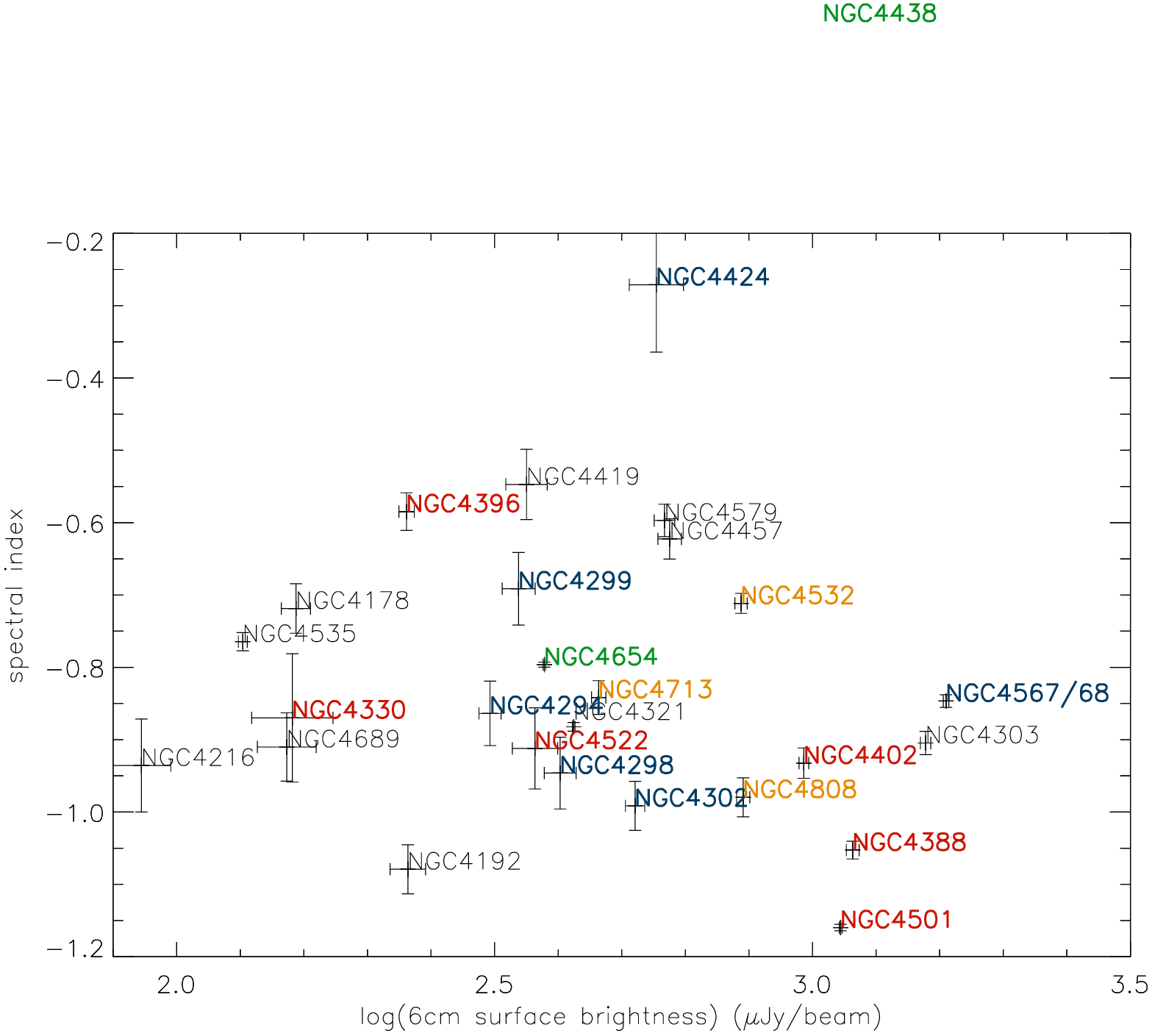}}
  \resizebox{10cm}{!}{\includegraphics{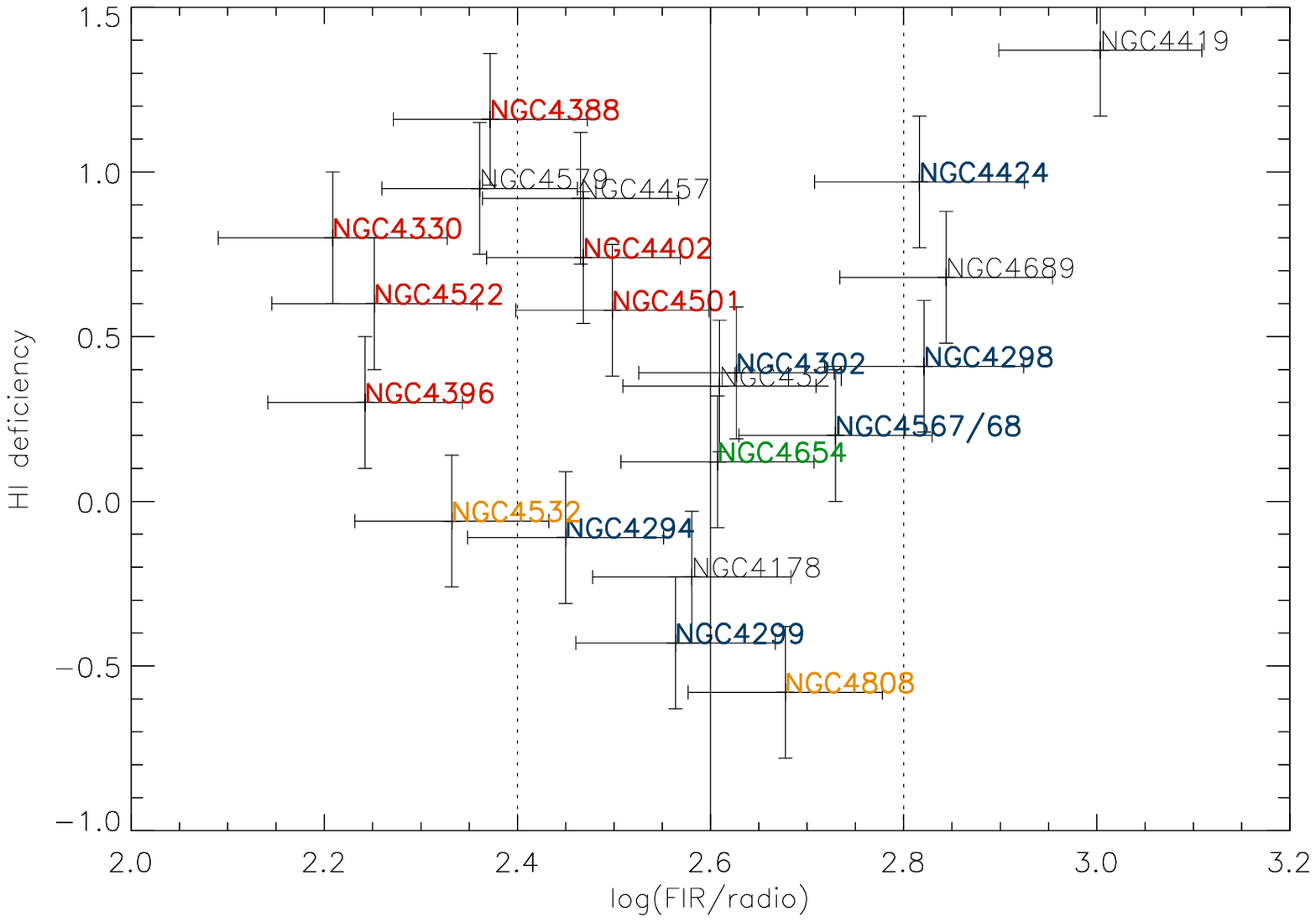}}
  \caption{Upper panel: average degree of polarization as a function of the cluster distance,
    middle panel: spectral index between 6 and 20~cm as a function of the mean 6cm total power surface brightness, lower panel:
    HI deficiency as a function of the FIR/radio ratio based on Spitzer $70$~$\mu$m flux densities. The vertical lines represent the mean and a scatter
    of 0.2~dex. Red: galaxies experiencing ram pressure stripping, blue: galaxies
    experiencing a tidal interaction, green: galaxies experiencing a tidal interaction and ram pressure
    stripping, orange: galaxies with H{\sc i} envelopes, black: non-interacting.
  \label{fig:stats2}}
\end{figure*}

To further investigate the radio continuum properties of our galaxy sample, the spectral index between 20~cm and 6~cm as a function of
the 6~cm mean total power surface brightness is presented in the middle panel of Fig.~\ref{fig:stats2}.
One might expect a steeper spectrum for a higher surface brightness, because synchrotron emission increases faster than thermal emission 
(Niklas \& Beck 1997). However, in agreement with Vollmer et al. (2004b) this trend is not observed in our galaxy sample
(Spearman rank coefficient of $-0.09$). 
NGC~4424 has a very shallow spectral index consistent with its classification as H{\sc ii} galaxy and the existence of a nuclear outflow (Sect.~\ref{sec:n4424}).
The spectral index of NGC~4438 is highly uncertain, because of the 20~cm image contamination with strong sidelobes from M~87 (Vollmer et al. 2009).

\begin{figure}
  \centering
  \resizebox{\hsize}{!}{\includegraphics{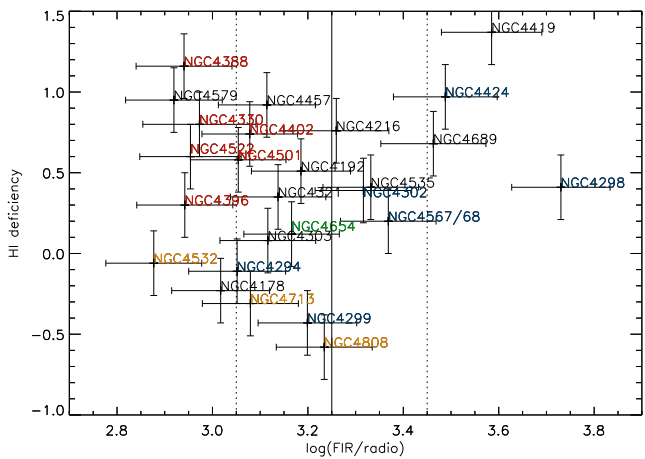}}
\caption{
    H{\sc i} deficiency as a function of the FIR/radio ratio based on IRAS $60$~$\mu$m and $100$~$\mu$m flux densities. 
    The vertical lines represent the mean and a scatter of 0.2~dex. 
    Red: galaxies experiencing ram pressure stripping, blue: galaxies
    experiencing a tidal interaction, green: galaxies experiencing a tidal interaction and ram pressure
    stripping, orange: galaxies with H{\sc i} envelopes, black: non-interacting.
  \label{fig:iras}}
\end{figure}

The nearly universal correlation between the far-infrared (FIR) and nonthermal radio continuum emission
of normal galaxies is known for several decades (e.g., de Jong et al. 1985; Helou et al. 1985).
The mean ratio between the 6~cm radio continuum and IRAS 60~$\mu$m flux densities is $\langle \log(S_{60\mu{\rm m}}/S_{\rm 6cm}) \rangle =2.4$
with a scatter of $\sim 0.2$ (de Jong et al. 1985, Wunderlich \& Klein 1988, We\.zgowiec et al. 2007).
Transforming the IRAS 60~$\mu$m flux densities into Spitzer 70~$\mu$m flux densities yields $S_{70\mu{\rm m}} \sim 1.5 \times S_{60\mu{\rm m}}$
(Murphy et al. 2006). We thus obtain $\langle \log(S_{70\mu{\rm m}}/S_{\rm 6cm}) \rangle = 2.6$.
The FIR/radio ratio of our 22 sample galaxies with available Spitzer $70$~$\mu$m flux densities is shown in the lower panel of Fig.~\ref{fig:stats2} 
together with this mean value and a scatter of $0.2$. For comparison, we also show the FIR/radio ratio based on the IRAS $60$~$\mu$m and $100$~$\mu$m flux densities
in Fig.~\ref{fig:iras} with the mean derived by Vollmer et al. (2004b) and a scatter of $0.2$~dex. 
Whereas the exact positions of the galaxies change in the diagrams if the Spitzer or IRAS FIR measurements are taken into account, 
their positions relative to the mean are robust.
The FIR/radio ratio of Virgo spiral galaxies (Fig.~\ref{fig:iras}) has a mean of $3.14$ and a scatter of $0.22$. Due to the presence of
$11$ galaxies with $2.9 \ge \log({\rm FIR/radio}) \ge 3.1$ the mean value is $0.06$~dex smaller than that of large samples. On the other hand,
the scatter is comparable to that of large samples. As Murphy et al. (2009) we find a radio excess in galaxies which are undergoing ram 
pressure stripping (NGC~4330, NGC~4388, NGC~4396, NGC~4522). All galaxies affected by ram pressure have FIR/radio ratios lower than the mean value.
A radio excess is also found in NGC~4532 and NGC~4713, which host a large accreting H{\sc i} envelopes. A global radio deficit is found in the tidally interacting 
galaxy NGC~4298, the H{\sc ii} galaxy NGC~4424, and NGC~4689. Surprisingly, NGC~4579 shows a radio excess,
whereas NGC~4548, a spiral galaxy of similar mass, morphology (including a prominent bar), and low overall gas surface density has a global radio 
deficit (We\.zgowiec et al. 2007). 
 
We thus conclude that active ram pressure stripping has no influence on the spectral index, but
enhances the global radio continuum emission with respect to the FIR emission by up to a factor of two (see also Murphy et al. 2009), 
while an accreting gas envelope can but not necessarily enhances the radio continuum emission with respect to the FIR emission.

\section{Conclusions\label{sec:conclusions}}

Vollmer et al. (2010) presented deep 20~cm and 6~cm VLA imaging observations of 8 Virgo galaxies. Most of these
galaxies are undergoing active ram pressure stripping. We extended this work with deep 20~cm and 6~cm VLA observations
of another 19 Virgo galaxies (Table~\ref{tab:table}). The total power, polarized intensity, projected magnetic field, and
degree of polarization maps were compared to optical, H$\alpha$, and H{\sc i} maps.
The integrated galaxy properties (radio continuum, H{\sc i}, FIR) of the full sample ($8+19$) were analyzed with respect to
the inclination, rotation velocity, and projected distance from the cluster center.

We found that
\begin{enumerate}
\item
the local spectral index mostly depends local star formation activity (H$\alpha$) as in field galaxies.
Only very strong ram pressure might lead to a flattening of the spectral index at the windward side of the galactic disk (Vollmer et al. 2010). 
\item
In two galaxies (NGC~4298, NGC~4457) the radio continuum distribution extends further than the H{\sc i} distribution.
In contrast to similar findings of Vollmer et al. (2010) these galaxies are not seen edge-on. NGC~4298 is interacting
tidally with NGC~4302, NGC~4457 probably had a recent minor merger.
\item
Six galaxies show a global minimum of 20~cm polarized emission at the receding side of the galactic disk. 
According to Braun et al. (2010), five of these galaxies have a thick magnetic disk (several kpc) with a quadrupole topology.
\item
Three galaxies (NGC~4303, NGC~4579, NGC~4713) display very smooth, symmetric spiral patterns. In NGC~4303 and NGC~4579 the 6~cm polarized emission
is even exceptionally smooth and not concentrated in the interarm regions as in many field spiral galaxies.
\item
Almost all highly inclined spiral galaxies, except NGC~4216, host a radio continuum halo. 
The projected magnetic field vectors of most of the symmetric halos show a characteristic X structure (NGC~4192,
NGC~4302, NGC~4419, NGC~4532).
\item
A galaxy with a strongly truncated gas disk (NGC~4419) can develop an extended radio continuum halo with an X structure of the magnetic field.
\item
Five galaxies of the extended sample of 27 Virgo galaxies show an asymmetric radio continuum halo:
NGC~4330, NGC~4457, NGC~4396, NGC~4402, NGC~4438.
Except NGC~4457, which probably had a recent minor merger, all galaxies are located in projection within the inner $3^{\circ}$(0.9~Mpc) of the cluster
and are undergoing ram pressure stripping.
\item 
Six out of 19 galaxies (NGC~4294, NGC~4298, NGC~4457, NGC~4532, NGC~4568, NGC~4808) display asymmetric 6~cm polarized emission distributions.
Three galaxies belong to tidally interacting pairs (NGC~4294, NGC~4298, NGC~4568), two galaxies host huge accreting H{\sc i} envelopes (NGC~4532, NGC~4808),
and one galaxy (NGC~4457) had a recent minor merger.
\item
In four out of the 6 galaxies with asymmetric 6~cm polarized emission distributions  the magnetic field is dominated by a component vertical to
the galactic disk (NGC~4294, NGC~4568, NGC~4532, NGC~4808). It seems that the polarized emission of the thin disk is very weak in these galaxies.
\item
The 6~cm average degree of polarization only correlates with the galaxy mass or rotation velocity. In addition, galaxies with low average star
formation rate per unit area have a low average degree of polarization. Shear or compression motions can enhance the degree of polarization.
The average degree of polarization of tidally interacting galaxies is generally lower than expected for a given rotation velocity and
star formation activity. The average degree of polarization of ram-pressure stripped galaxies can but has not to be lower than expected.
\item
Virgo spiral galaxies undergoing ram pressure stripping show a global radio excess with respect to their FIR emission as already
shown be Murphy et al. (2009). A  radio excess is also found in NGC~4532 and NGC~4713, which host large accreting H{\sc i} envelopes, and the massive,
highly H{\sc i}-deficient Virgo spiral galaxy NGC~4579.
\end{enumerate}

We conclude that
\begin{enumerate}
\item
In the new sample no additional case of a ram-pressure stripped spiral galaxy with an asymmetric ridge of polarized radio continuum
emission is found.
\item
In the absence of a close companion, a truncated H{\sc i} disk together with a ridge of polarized radio continuum emission at the
outer edge of the H{\sc i} disk are a signpost of active ram pressure stripping. The polarized radio ridge can be absent if ram pressure
compression occurs in the plane of the sky or erased by strong shear motions of the disk gas.
\item
Tidal interactions and accreting gas envelopes can also lead to compression and shear motions which enhance the polarized radio continuum emission.
The resulting asymmetries are located within the H{\sc i} distribution.
\item
Ram pressure stripping can decrease whereas tidal interactions most frequently decreases the average degree of 
polarization of Virgo cluster spiral galaxies. The low average degree of polarization in tidally interacting galaxies is
consistent with the absence of polarized emission from the thin disk.
\item
Not too strong active ram pressure stripping has no influence on the spectral index, but
enhances the global radio continuum emission with respect to the FIR emission by up to a factor of two (see also Murphy et al. 2009), 
while an accreting gas envelope can but not necessarily enhances the radio continuum emission with respect to the FIR emission.
\end{enumerate}

\begin{acknowledgements}
This research has made use of the GOLD Mine Database. This work was supported by the grant from National Science Centre 
in Poland (NCN) on the basis of decision number DEC-2011/03/B/ST9/01859.
RB acknowledges support by project DFG RU1254.
\end{acknowledgements}

\appendix

\section{Multiwavelenth maps of individual galaxies}

\begin{figure*}
  \centering
 \resizebox{15cm}{!}{\includegraphics{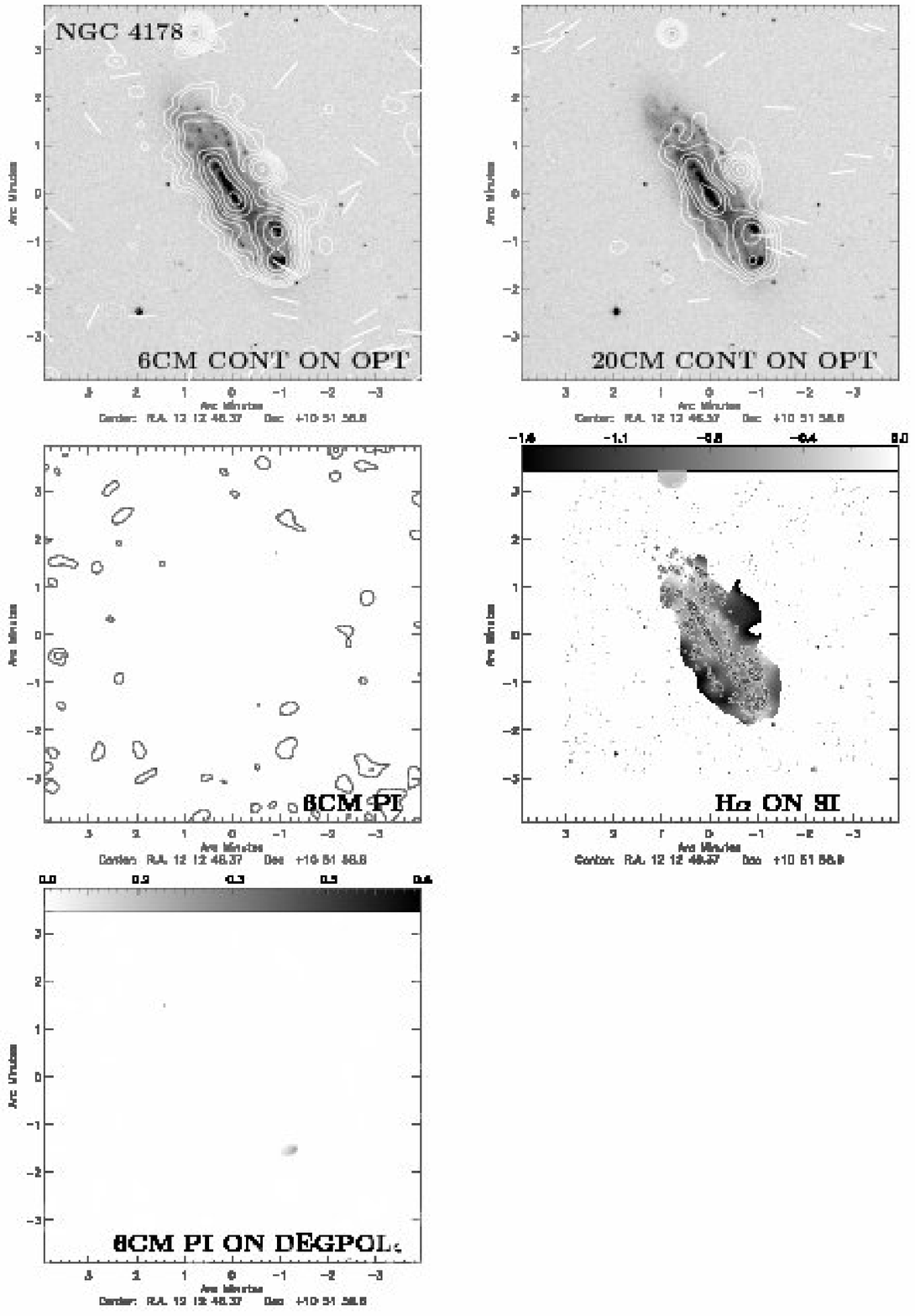}}
  \caption{NGC~4178: from top left to bottom right: 6~cm total power emission distribution on DSS B band image
    together with the apparent B vectors,
    20~cm total power emission distribution on DSS B band image together with the apparent B vectors,
    6~cm polarized emission distribution,
    Goldmine H$\alpha$ emission distribution on spectral index map, and 
    6~cm polarized emission distribution on degree of polarization.
    The sizes of the B vectors are proportional to the polarized intensity.
    Contour levels are $\xi \times (-3,3,5,8,12,20,30,50,80,120,200,300)$, with
    $\xi=12$~$\mu$Jy for the 6~cm total power emission, $\xi=80$~$\mu$Jy for the 20~cm total power emission, and
    $\xi=12$~$\mu$Jy for the 6~cm polarized emission.
    \label{fig:zusammen1n4178}}%
\end{figure*}

\begin{figure*}
  \centering
 \resizebox{15cm}{!}{\includegraphics{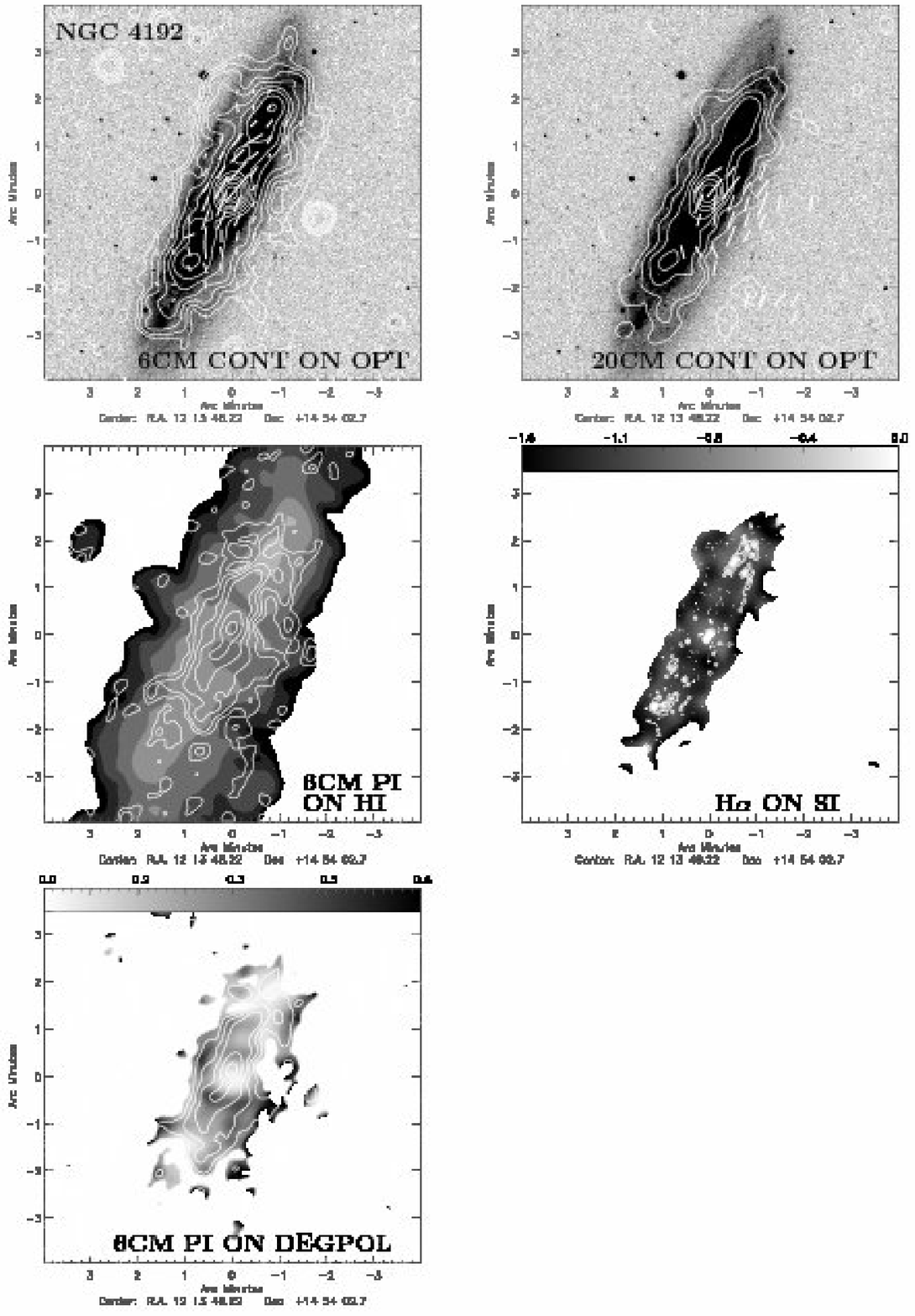}}
  \caption{NGC~4192: from top left to bottom right: 6~cm total power emission distribution on DSS B band image
    together with the apparent B vectors,
    20~cm total power emission distribution on DSS B band image together with the apparent B vectors,
    6~cm polarized emission distribution on VIVA H{\sc i} distribution,
    Goldmine H$\alpha$ emission distribution on spectral index map, and 
    6~cm polarized emission distribution on degree of polarization.
    The sizes of the B vectors are proportional to the polarized intensity.
    Contour levels are $\xi \times (-3,3,5,8,12,20,30,50,80,120,200,300)$, with
    $\xi=16$~$\mu$Jy for the 6~cm total power emission, $\xi=210$~$\mu$Jy for the 20~cm total power emission, and
    $\xi=10$~$\mu$Jy for the 6~cm polarized emission.
     \label{fig:zusammen1n4192}}%
\end{figure*}

\begin{figure*}
  \centering
 \resizebox{15cm}{!}{\includegraphics{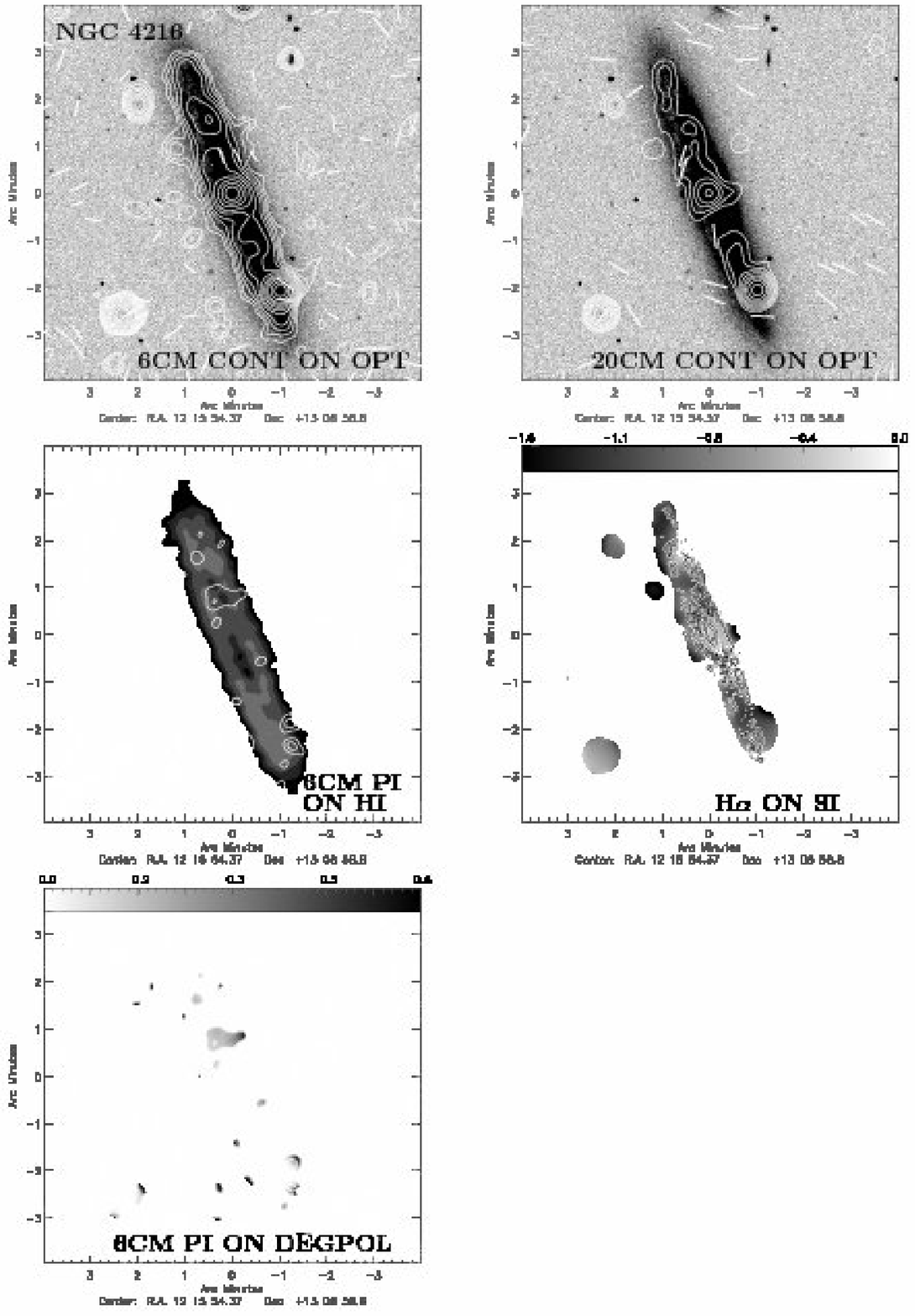}}
  \caption{NGC~4216: same panels as in Fig.~\ref{fig:zusammen1n4192}.
    Contour levels are $\xi \times (-3,3,5,8,12,20,30,50,80,120,200,300)$, with
    $\xi=9$~$\mu$Jy for the 6~cm total power emission, $\xi=110$~$\mu$Jy for the 20~cm total power emission, and 
    $\xi=7$~$\mu$Jy for the 6~cm polarized emission.
     \label{fig:zusammen1n4216}}%
\end{figure*}

\begin{figure*}
  \centering
 \resizebox{15cm}{!}{\includegraphics{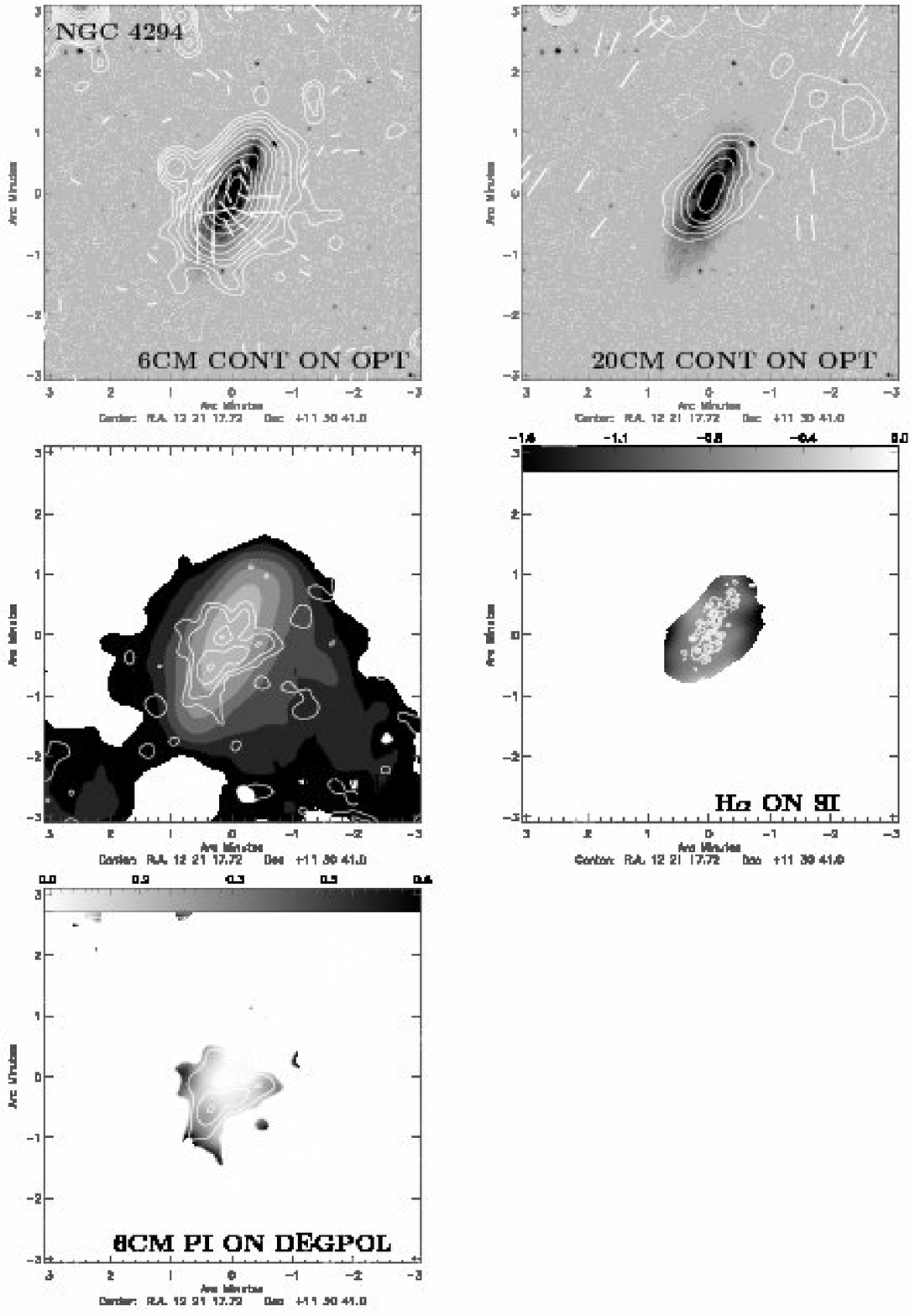}}
  \caption{NGC~4294: same panels as in Fig.~\ref{fig:zusammen1n4192}.
    Contour levels are $\xi \times (-3,3,5,8,12,20,30,50,80,120,200,300)$, with
    $\xi=10$~$\mu$Jy for the 6~cm total power emission, $\xi=200$~$\mu$Jy for the 20~cm total power emission, and
    $\xi=8$~$\mu$Jy for the 6~cm polarized emission.
     \label{fig:zusammen1n4294}}%
\end{figure*}

\clearpage

\begin{figure*}
  \centering
 \resizebox{15cm}{!}{\includegraphics{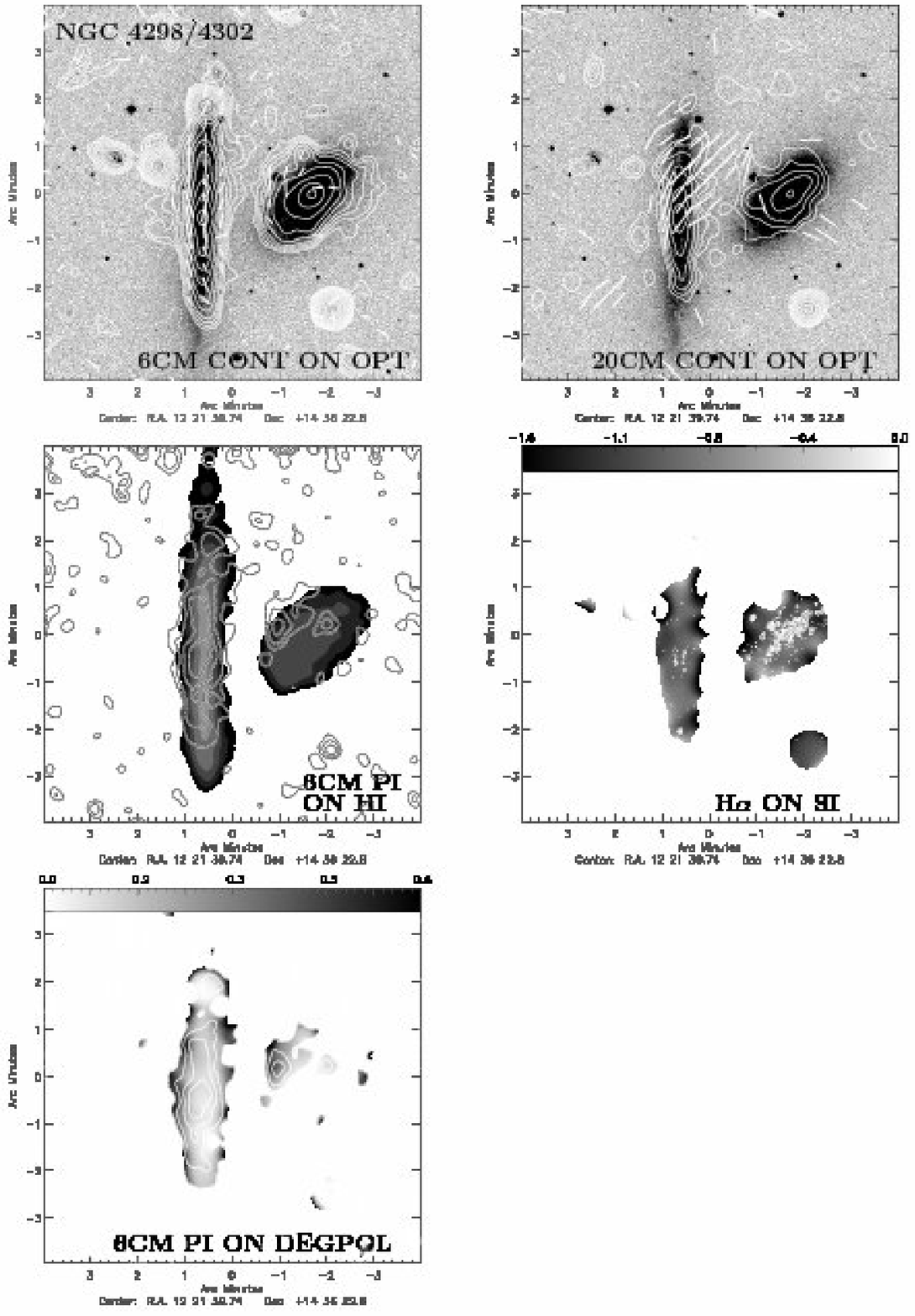}}
  \caption{NGC~4298 (western galaxy): same panels as in Fig.~\ref{fig:zusammen1n4192}.
    Contour levels are $\xi \times (-3,3,5,8,12,20,30,50,80,120,200,300)$, with
    $\xi=11$~$\mu$Jy for the 6~cm total power emission, $\xi=170$~$\mu$Jy for the 20~cm total power emission, and
    $\xi=10$~$\mu$Jy for the 6~cm polarized emission.
     \label{fig:zusammen1n4298}}%
\end{figure*}

\begin{figure*}
  \centering
 \resizebox{15cm}{!}{\includegraphics{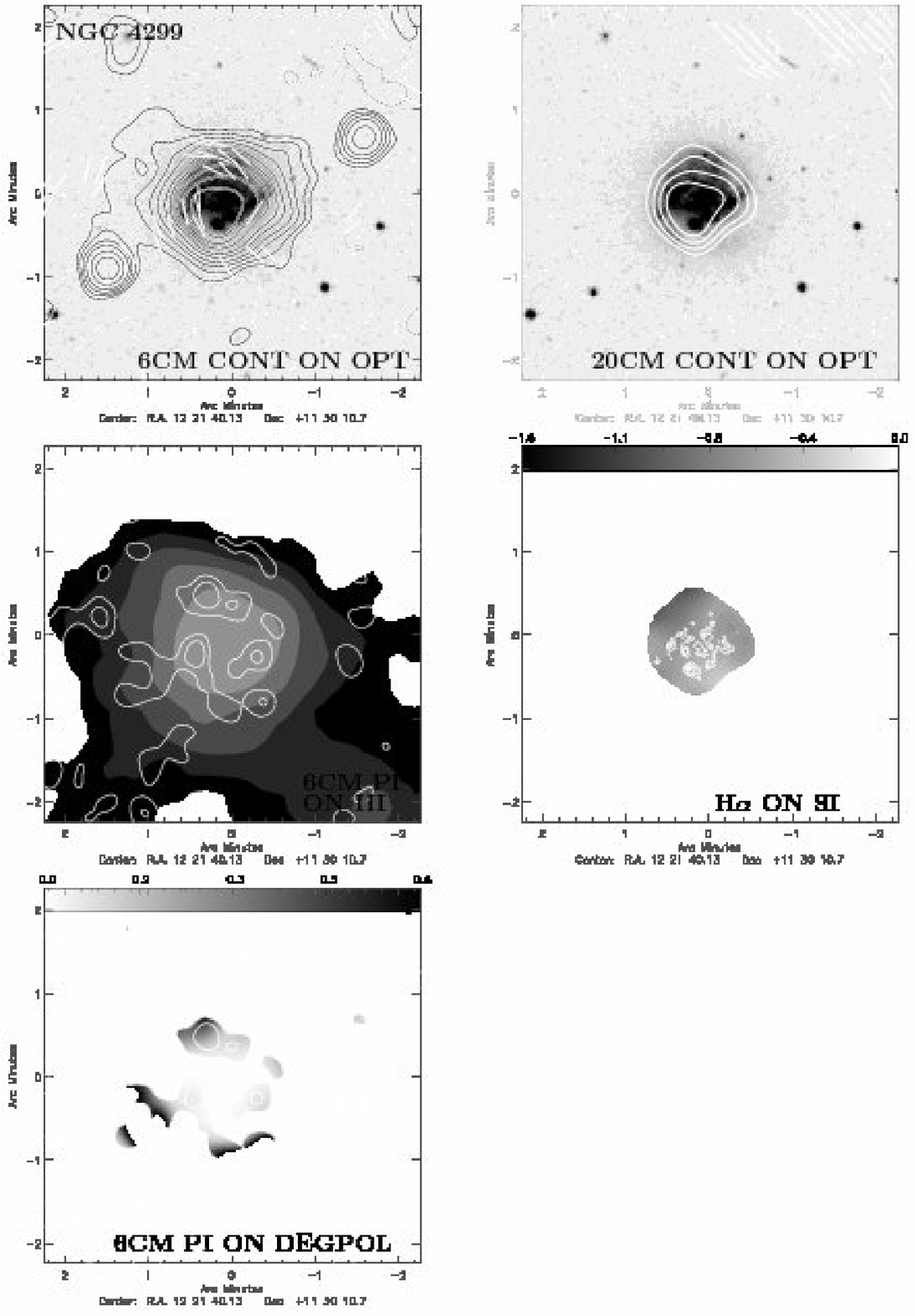}}
  \caption{NGC~4299: same panels as in Fig.~\ref{fig:zusammen1n4192}.
    Contour levels are $\xi \times (-3,3,5,8,12,20,30,50,80,120,200,300)$, with
    $\xi=10$~$\mu$Jy for the 6~cm total power emission, $\xi=180$~$\mu$Jy for the 20~cm total power emission, and
    $\xi=8$~$\mu$Jy for the 6~cm polarized emission.
     \label{fig:zusammen1n4299}}%
\end{figure*}

\begin{figure*}
  \centering
 \resizebox{15cm}{!}{\includegraphics{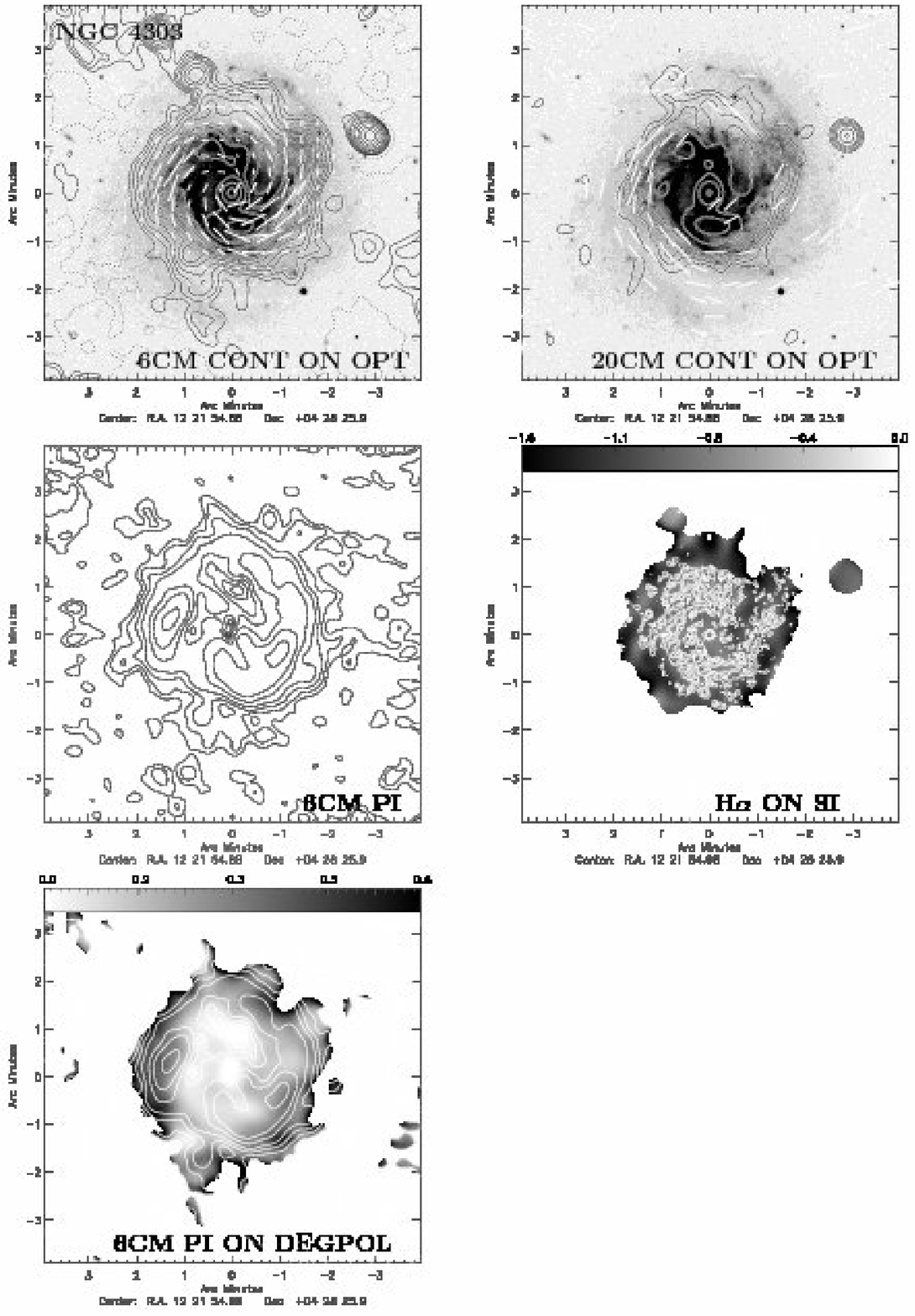}}
  \caption{NGC~4303: same panels as in Fig.~\ref{fig:zusammen1n4178}.
    Contour levels are $\xi \times (-3,3,5,8,12,20,30,50,80,120,200,300)$, with
    $\xi=30$~$\mu$Jy for the 6~cm total power emission, $\xi=450$~$\mu$Jy for the 20~cm total power emission, and
    $\xi=12$~$\mu$Jy for the 6~cm polarized emission.
     \label{fig:zusammen1n4303}}%
\end{figure*}

\begin{figure*}
  \centering
 \resizebox{15cm}{!}{\includegraphics{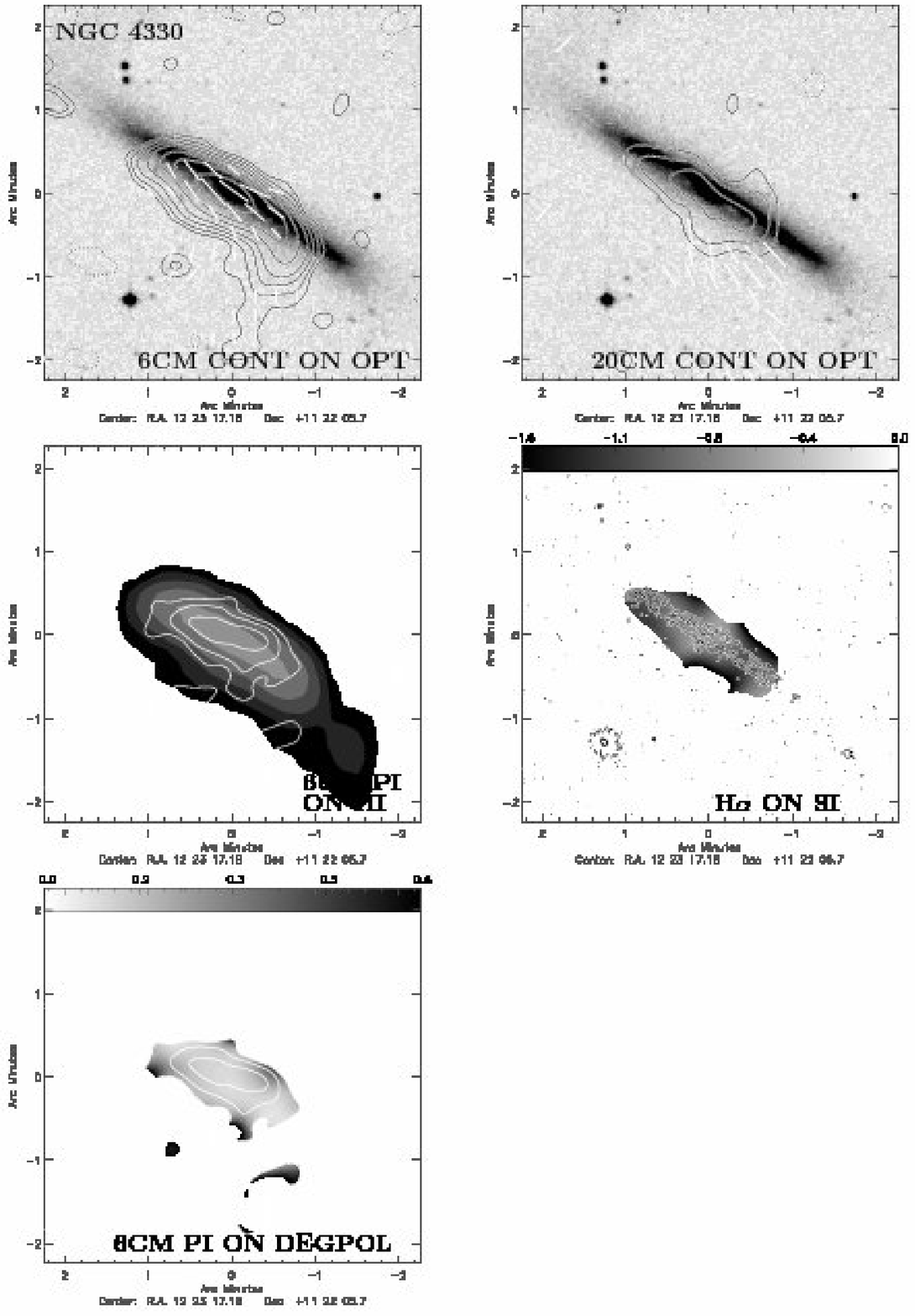}}
  \caption{NGC~4330: same panels as in Fig.~\ref{fig:zusammen1n4192}.
    Contour levels are $\xi \times (-3,3,5,8,12,20,30,50,80,120,200,300)$, with
    $\xi=14$~$\mu$Jy for the 6~cm total power emission, $\xi=210$~$\mu$Jy for the 20~cm total power emission, and
    $\xi=10$~$\mu$Jy for the 6~cm polarized emission.
     \label{fig:zusammen1n4330}}%
\end{figure*}

\clearpage

\begin{figure*}
  \centering
 \resizebox{15cm}{!}{\includegraphics{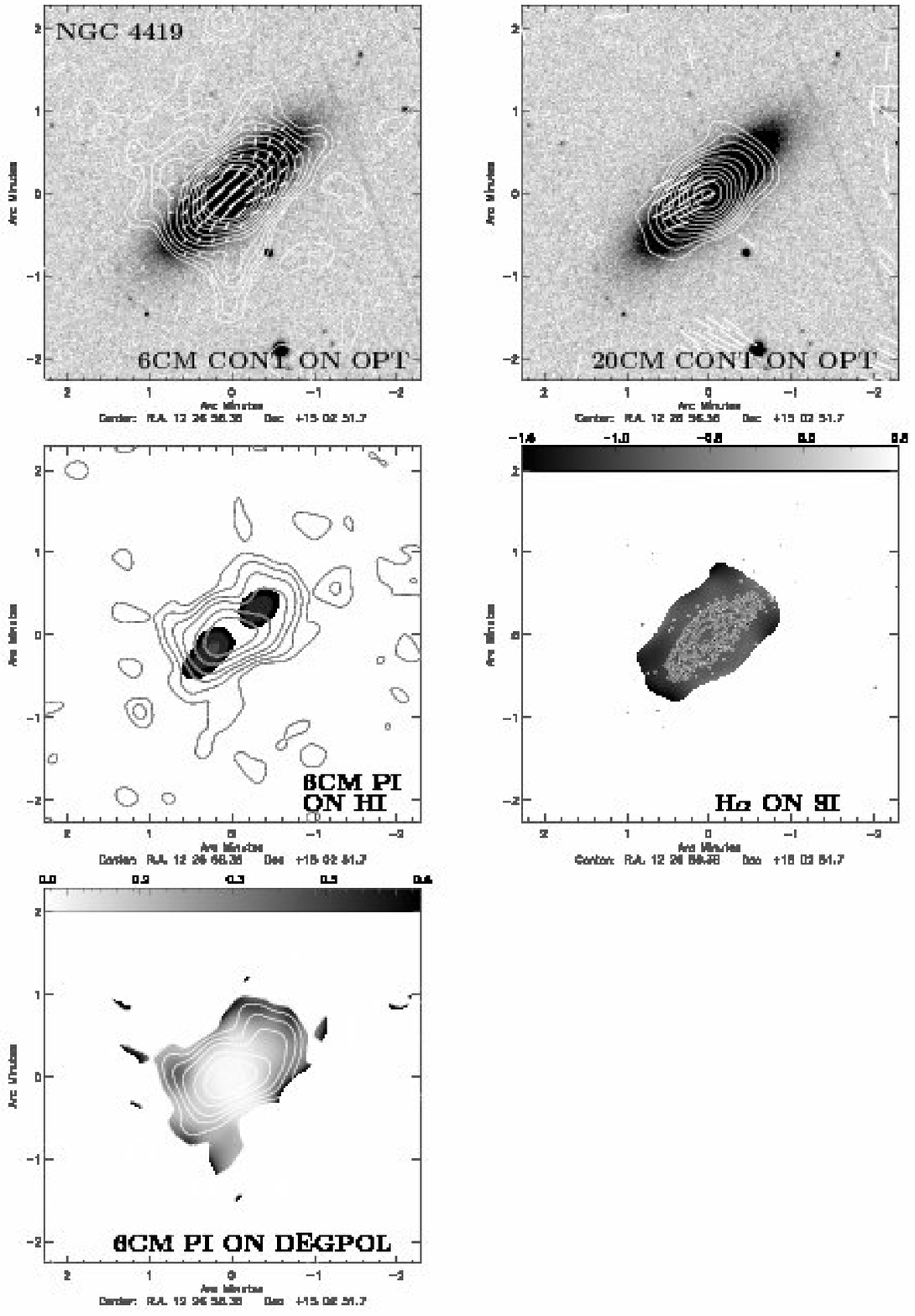}}
  \caption{NGC~4419: same panels as in Fig.~\ref{fig:zusammen1n4192}.
    Contour levels are $\xi \times (-3,3,5,8,12,20,30,50,80,120,200,300)$, with
    $\xi=11$~$\mu$Jy for the 6~cm total power emission, $\xi=130$~$\mu$Jy for the 20~cm total power emission, and
    $\xi=7$~$\mu$Jy for the 6~cm polarized emission.
     \label{fig:zusammen1n4419}}%
\end{figure*}

\begin{figure*}
  \centering
 \resizebox{15cm}{!}{\includegraphics{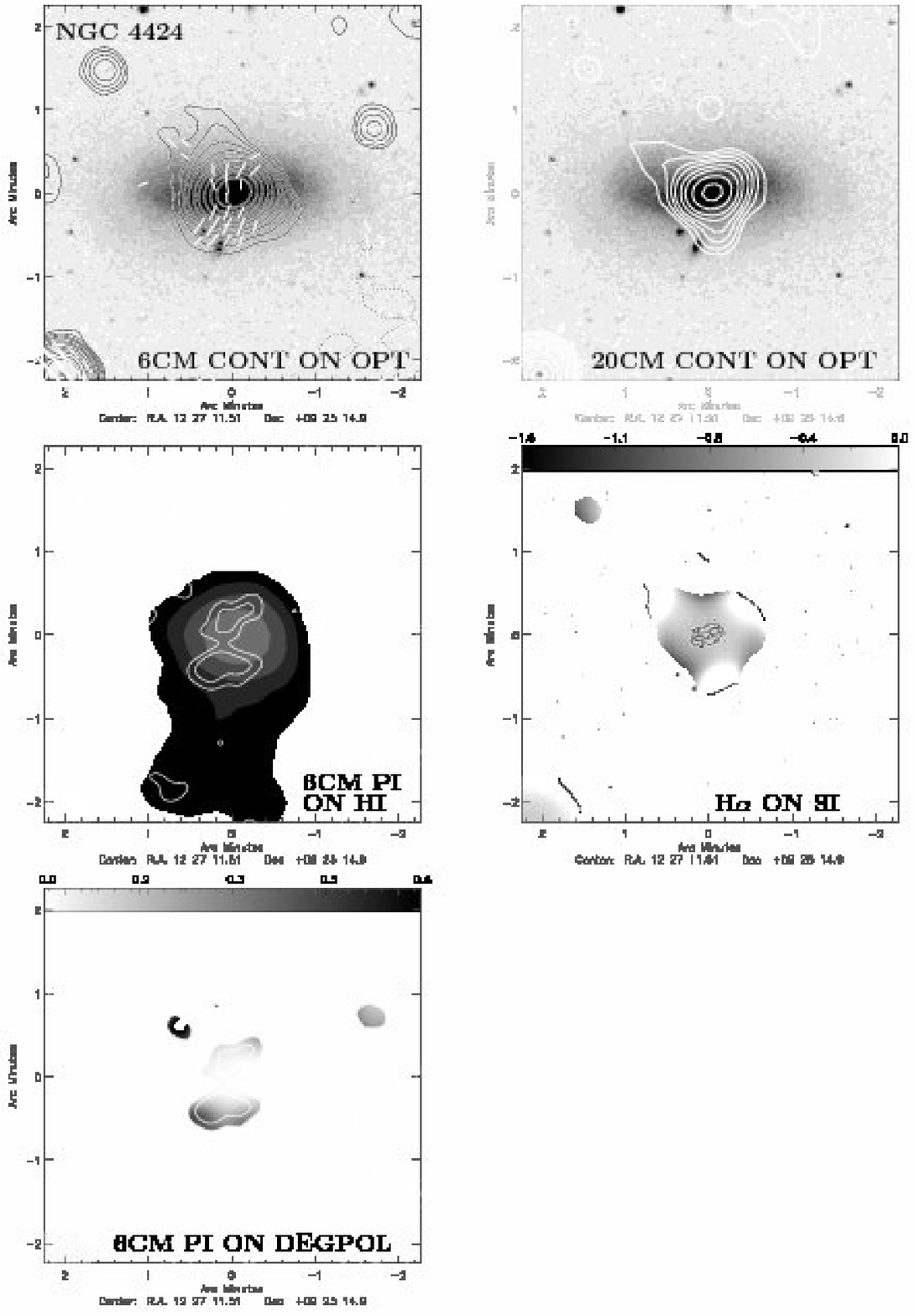}}
  \caption{NGC~4424: same panels as in Fig.~\ref{fig:zusammen1n4192}.
    Contour levels are $\xi \times (-3,3,5,8,12,20,30,50,80,120,200,300)$, with
    $\xi=14$~$\mu$Jy for the 6~cm total power emission, $\xi=20$~$\mu$Jy for the 20~cm total power emission,
    and $\xi=7$~$\mu$Jy for the 6~cm polarized emission.
     \label{fig:zusammen1n4424}}%
\end{figure*}

\begin{figure*}
  \centering
 \resizebox{15cm}{!}{\includegraphics{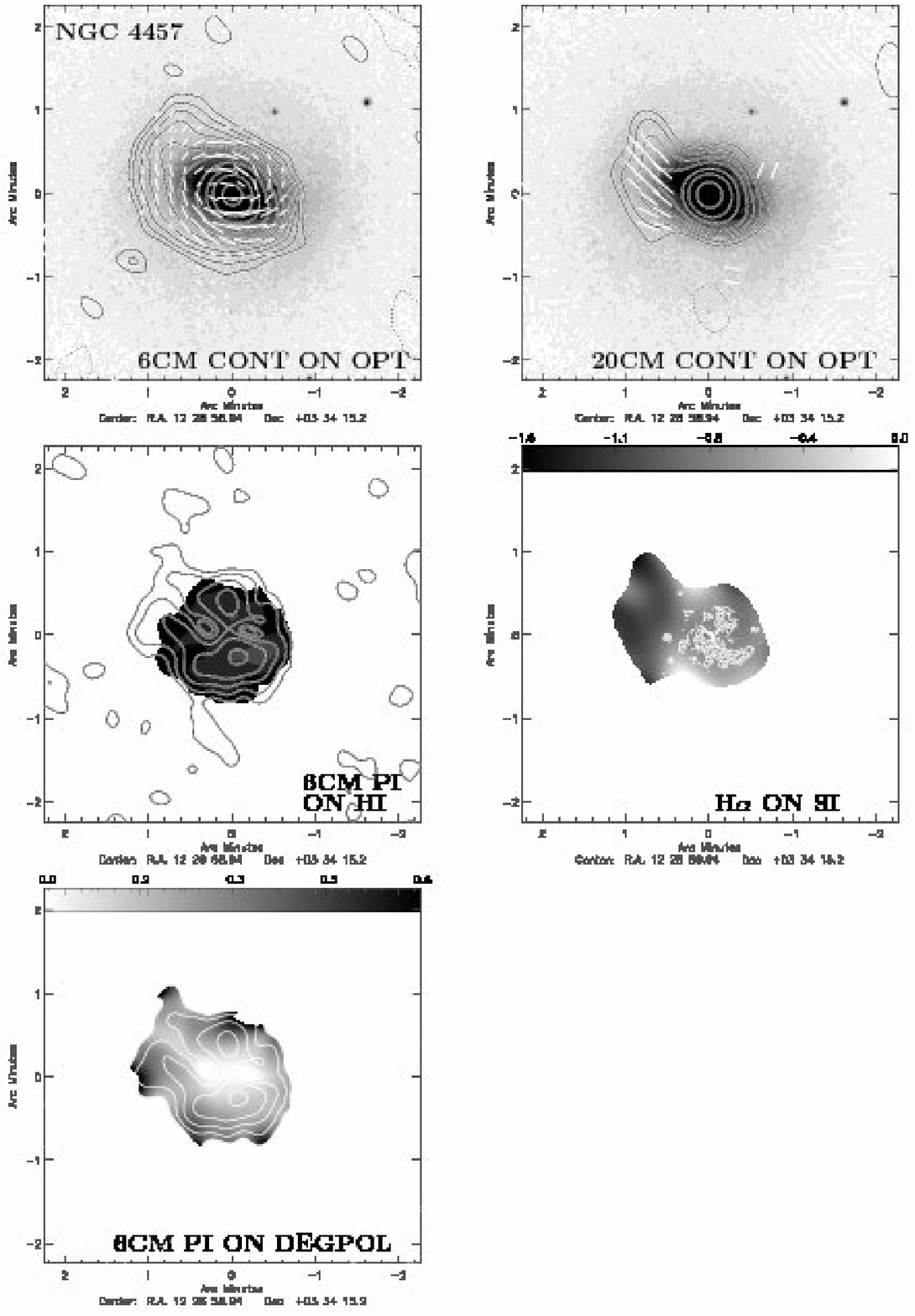}}
  \caption{NGC~4457: same panels as in Fig.~\ref{fig:zusammen1n4192}.
    Contour levels are $\xi \times (-3,3,5,8,12,20,30,50,80,120,200,300)$, with
    $\xi=23$~$\mu$Jy for the 6~cm total power emission, $\xi=190$~$\mu$Jy for the 20~cm total power emission,
    and $\xi=15$~$\mu$Jy for the 6~cm polarized emission.
     \label{fig:zusammen1n4457}}%
\end{figure*}

\begin{figure*}
  \centering
 \resizebox{15cm}{!}{\includegraphics{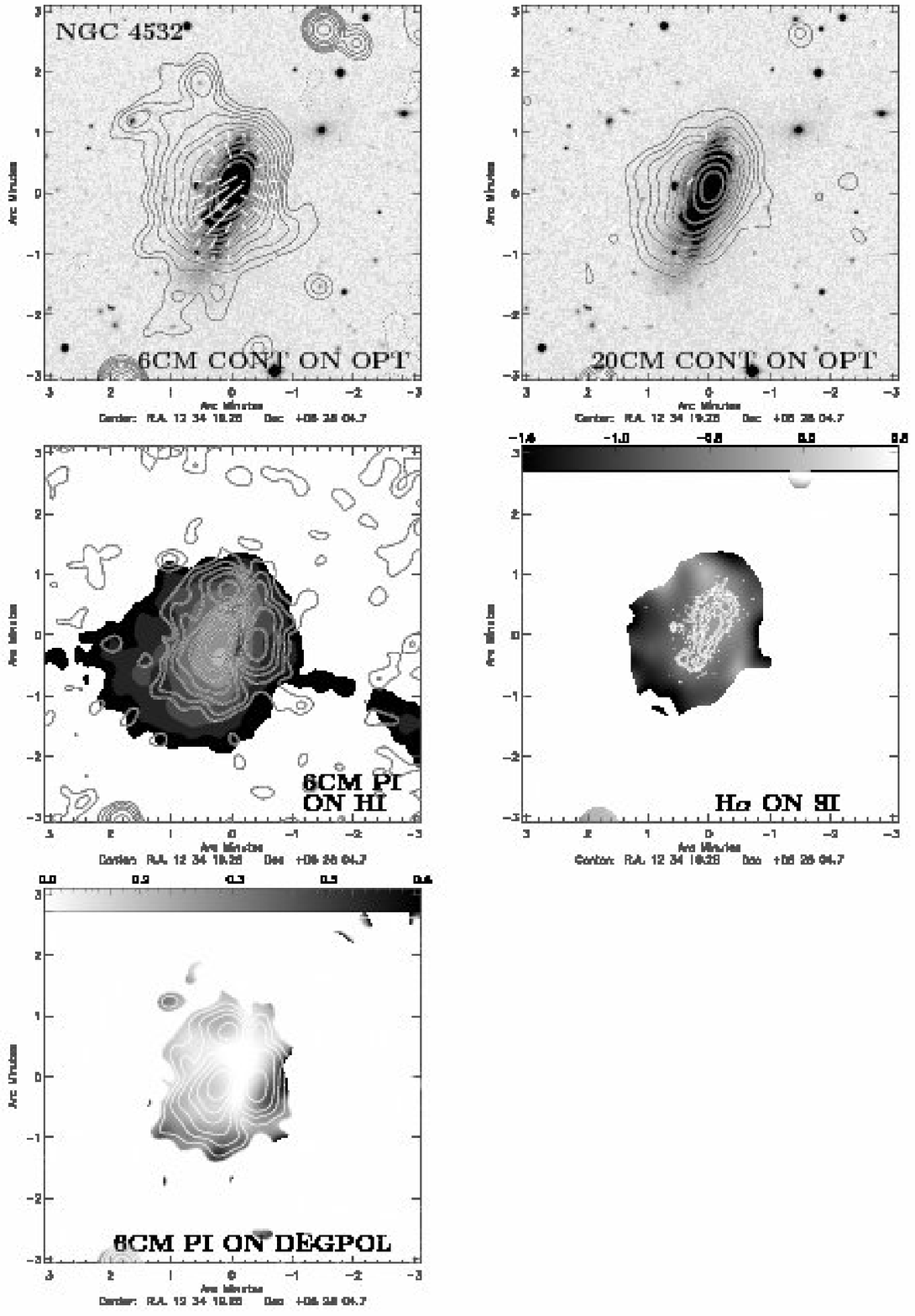}}
  \caption{NGC~4532: same panels as in Fig.~\ref{fig:zusammen1n4192}.
    Contour levels are $\xi \times (-3,3,5,8,12,20,30,50,80,120,200,300)$, with
    $\xi=18$~$\mu$Jy for the 6~cm total power emission, $\xi=160$~$\mu$Jy for the 20~cm total power emission,
    and $\xi=8$~$\mu$Jy for the 6~cm polarized emission.
     \label{fig:zusammen1n4532}}%
\end{figure*}

\clearpage

\begin{figure*}
  \centering
 \resizebox{15cm}{!}{\includegraphics{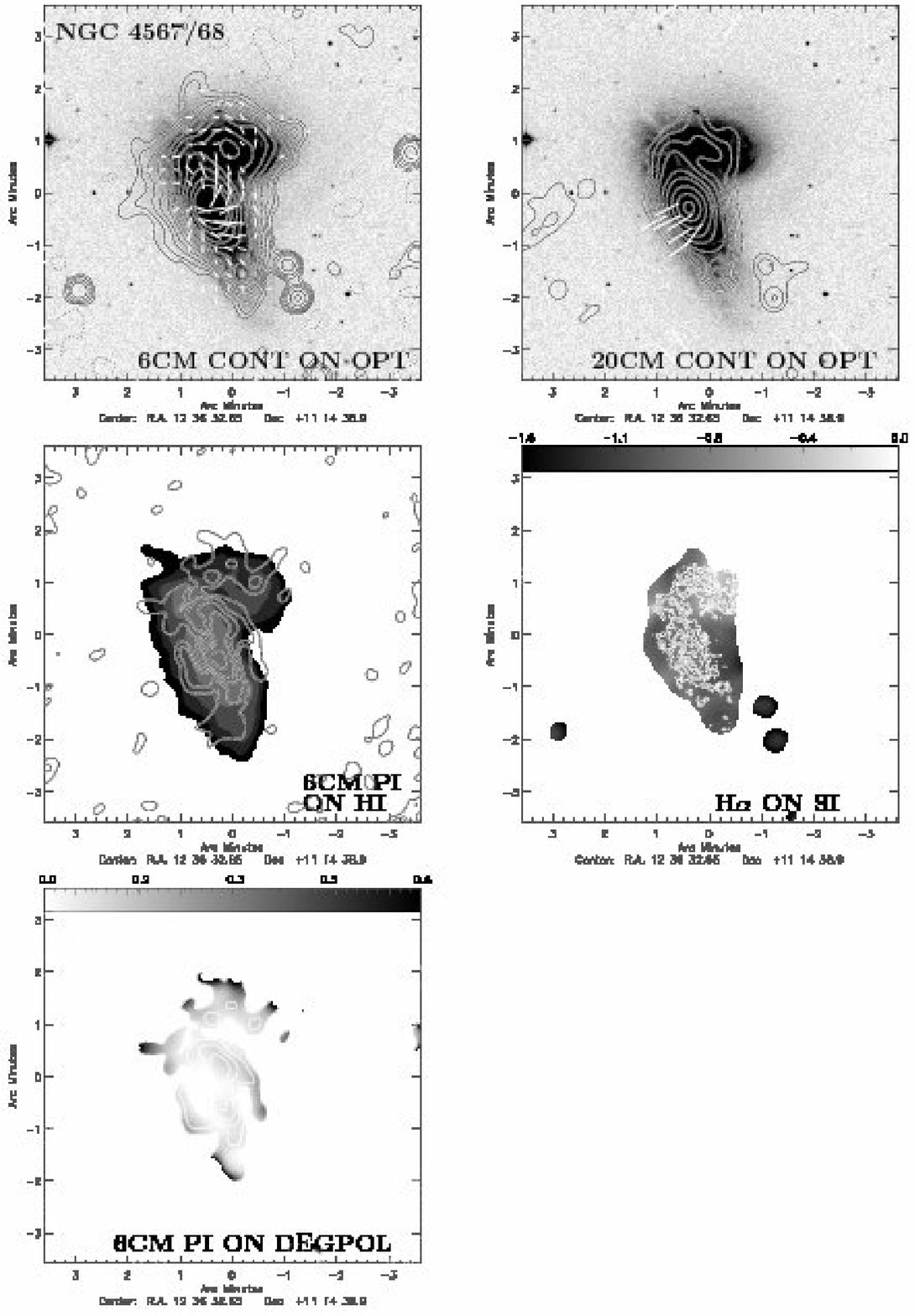}}
  \caption{NGC~4567/68: same panels as in Fig.~\ref{fig:zusammen1n4192}.
    Contour levels are $\xi \times (-3,3,5,8,12,20,30,50,80,120,200,300)$, with
    $\xi=14$~$\mu$Jy for the 6~cm total power emission, $\xi=230$~$\mu$Jy for the 20~cm total power emission,
    and $\xi=12$~$\mu$Jy for the 6~cm polarized emission.
     \label{fig:zusammen1n4567}}%
\end{figure*}

\begin{figure*}
  \centering
 \resizebox{15cm}{!}{\includegraphics{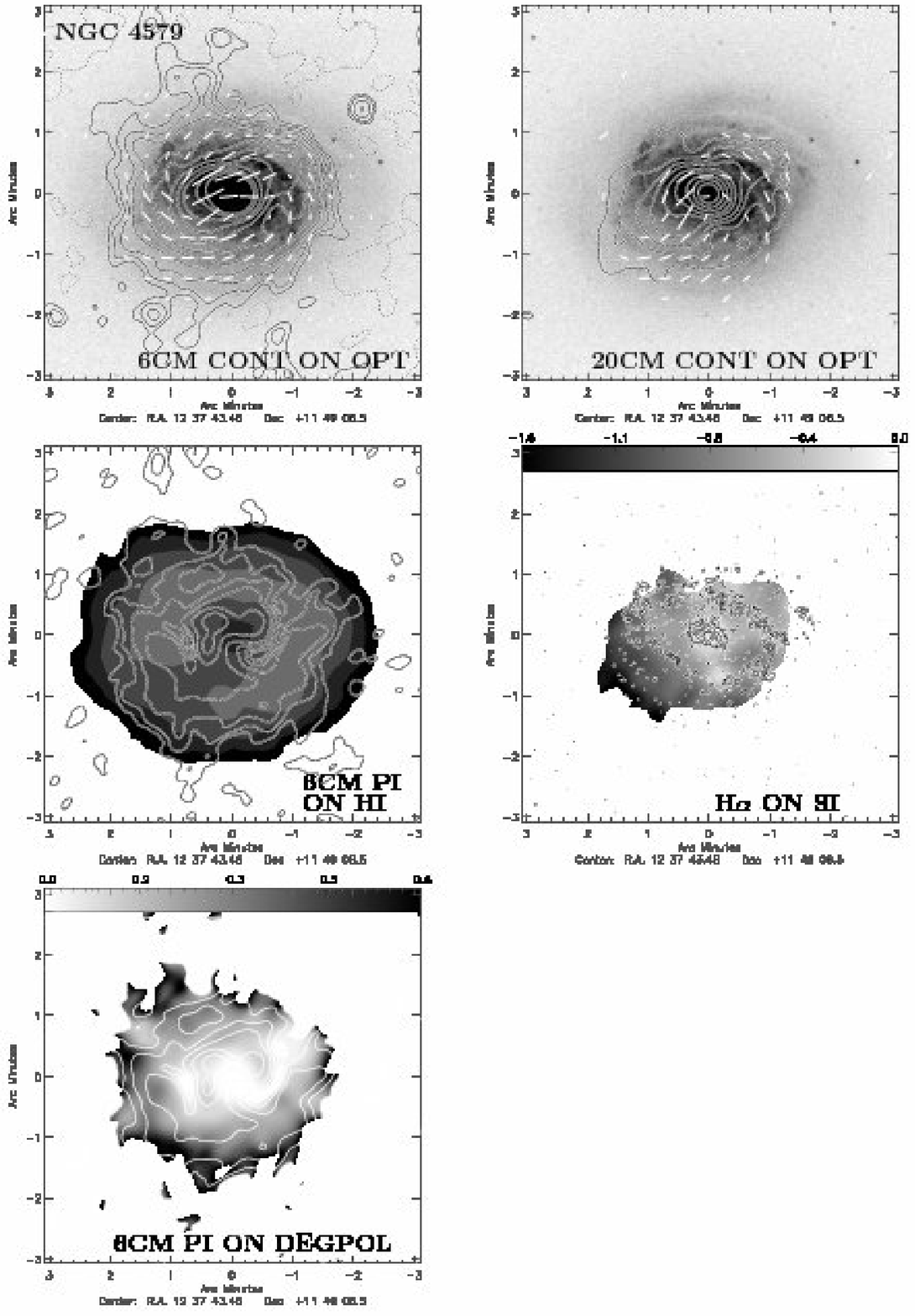}}
  \caption{NGC~4579: same panels as in Fig.~\ref{fig:zusammen1n4192}.
    Contour levels are $\xi \times (-3,3,5,8,12,20,30,50,80,120,200,300)$, with
    $\xi=18$~$\mu$Jy for the 6~cm total power emission, $\xi=250$~$\mu$Jy for the 20~cm total power emission,
    and $\xi=10$~$\mu$Jy for the 6~cm polarized emission.
     \label{fig:zusammen1n4579}}%
\end{figure*}

\begin{figure*}
  \centering
 \resizebox{15cm}{!}{\includegraphics{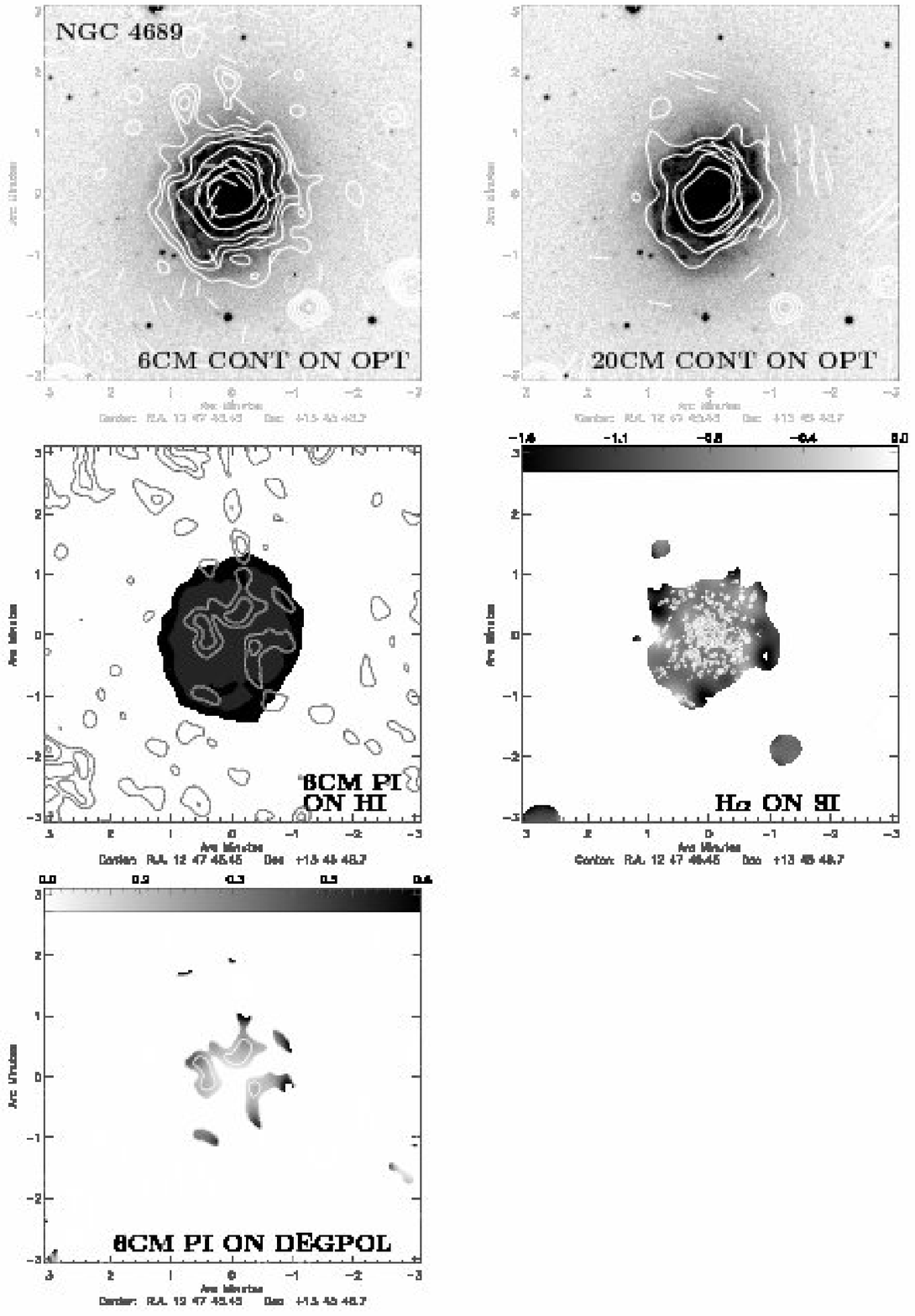}}
  \caption{NGC~4689:  same panels as in Fig.~\ref{fig:zusammen1n4192}.
    Contour levels are $\xi \times (-3,3,5,8,12,20,30,50,80,120,200,300)$, with
    $\xi=14$~$\mu$Jy for the 6~cm total power emission, $\xi=80$~$\mu$Jy for the 20~cm total power emission,
    and $\xi=12$~$\mu$Jy for the 6~cm polarized emission.
     \label{fig:zusammen1n4689}}%
\end{figure*}

\begin{figure*}
  \centering
 \resizebox{15cm}{!}{\includegraphics{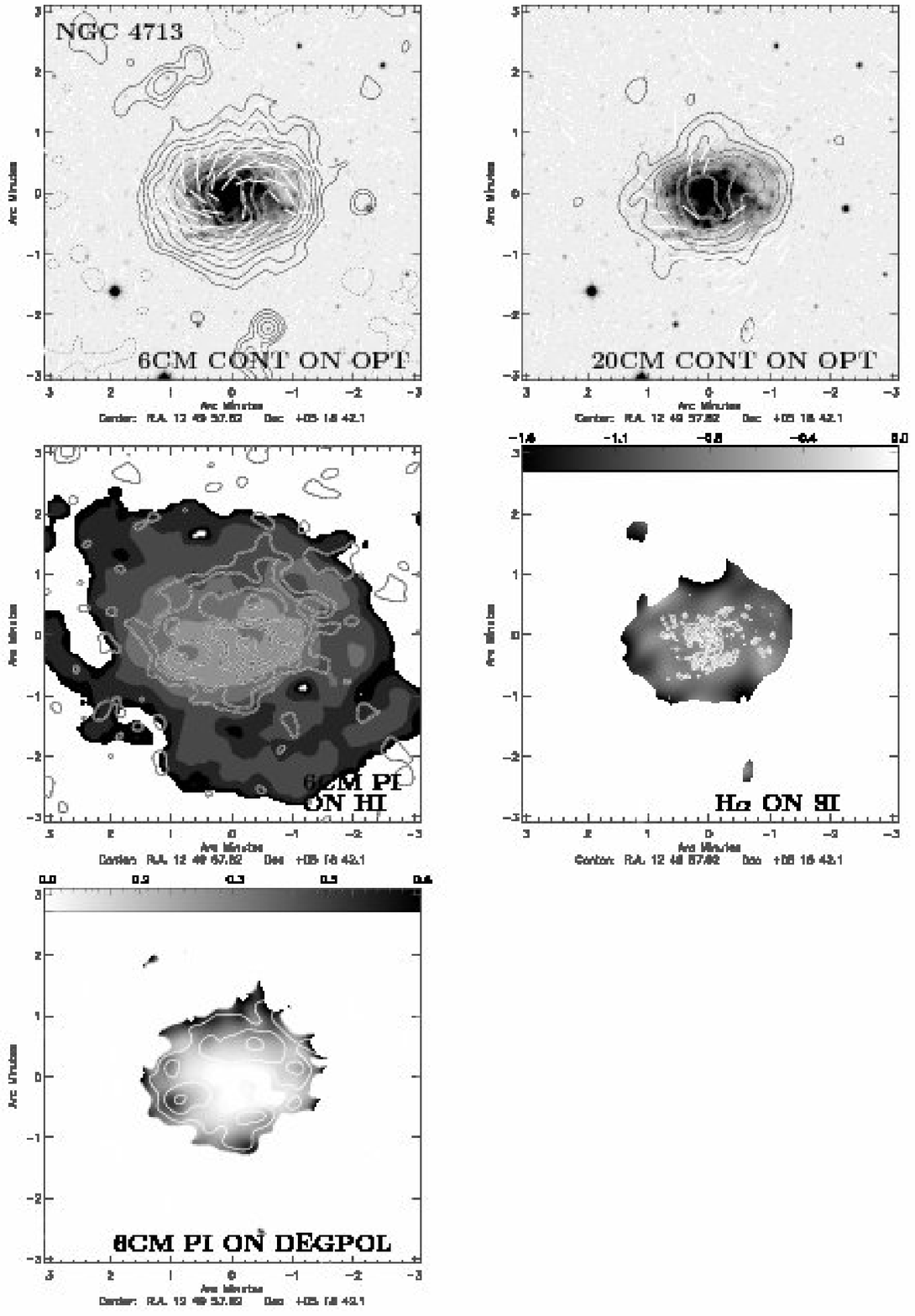}}
  \caption{NGC~4713: same panels as in Fig.~\ref{fig:zusammen1n4192}.
    Contour levels are $\xi \times (-3,3,5,8,12,20,30,50,80,120,200,300)$, with
    $\xi=12$~$\mu$Jy for the 6~cm total power emission, $\xi=120$~$\mu$Jy for the 20~cm total power emission,
    and $\xi=8$~$\mu$Jy for the 6~cm polarized emission.
     \label{fig:zusammen1n4713}}%
\end{figure*}

\clearpage

\begin{figure*}
  \centering
 \resizebox{15cm}{!}{\includegraphics{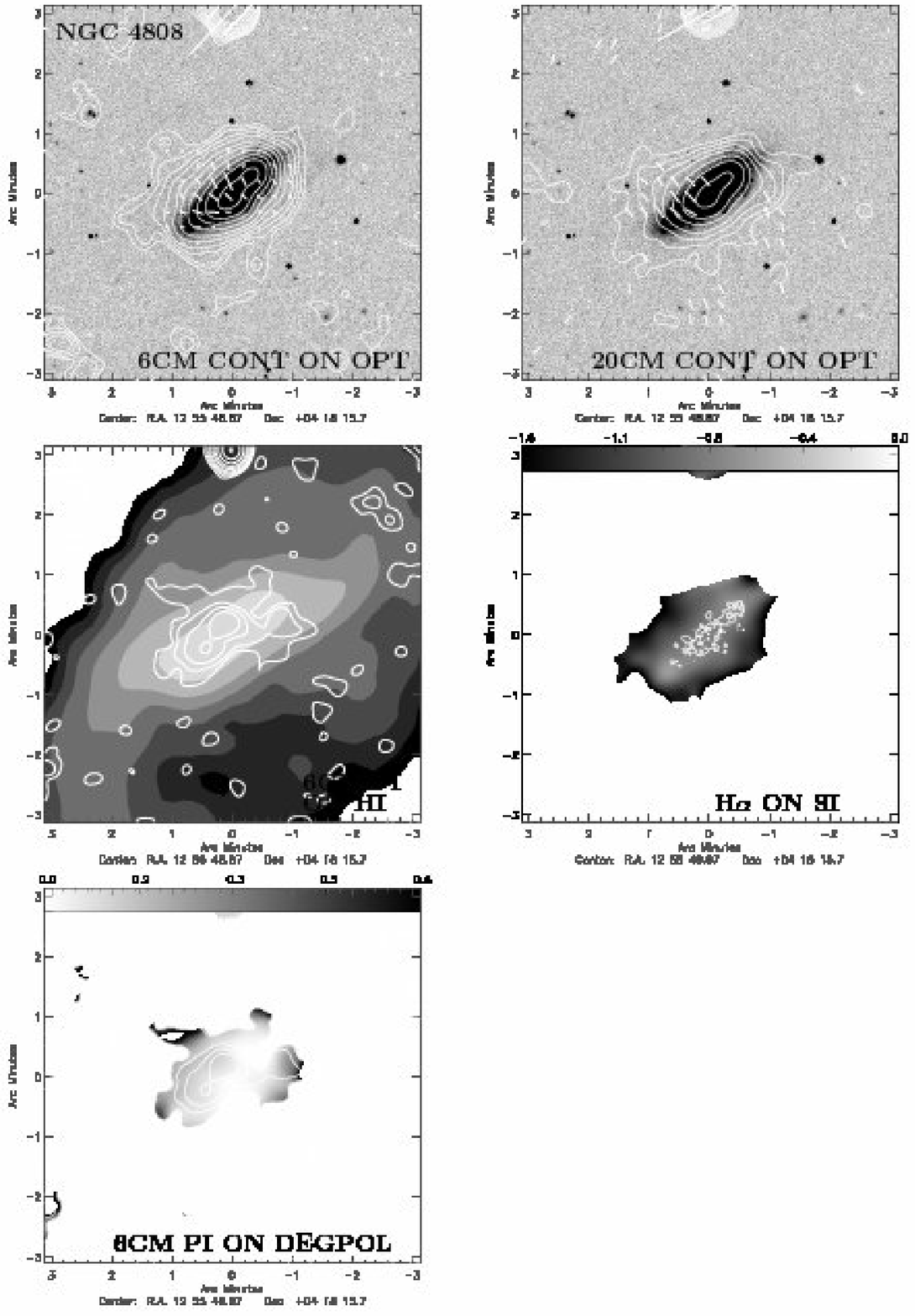}}
  \caption{NGC~4808: same panels as in Fig.~\ref{fig:zusammen1n4192}.
    Contour levels are $\xi \times (-3,3,5,8,12,20,30,50,80,120,200,300)$, with
    $\xi=14$~$\mu$Jy for the 6~cm total power emission, $\xi=140$~$\mu$Jy for the 20~cm total power emission,
    and $\xi=9$~$\mu$Jy for the 6~cm polarized emission.
     \label{fig:zusammen1n4808}}%
\end{figure*}

\end{document}